\documentclass[a4paper]{jpconf}
\usepackage{graphicx}
\usepackage{wrapfig}
\begin{document}
\title{Towards A Background Independent Quantum Gravity}

\author{Hyun Seok Yang}

\address{Institute for the Early Universe, Ewha Womans University, Seoul 120-750, Korea}

\ead{hsyang@sogang.ac.kr}

\begin{abstract}
We recapitulate the scheme of emergent gravity to highlight how a
background independent quantum gravity can be defined by quantizing
spacetime itself.
\end{abstract}

\section{Introduction}

According to general relativity, gravity is the dynamics of
spacetime geometry where spacetime is a (pseudo-)Riemannian manifold
and the gravitational field is represented by a Riemannian metric
\cite{big-book}. The gravitational field equations are given by the Einstein
equations defined by
\begin{equation}\label{einstein-eq}
    R_{MN} - \frac{1}{2} g_{MN} R = \frac{8 \pi G}{c^4} T_{MN}.
\end{equation}
A beauty of the equation (\ref{einstein-eq}) may be phrased in a
poetic diction (John A. Wheeler) that matter tells spacetime how to
curve, and spacetime tells matter how to move. However the notorious
difficulty to elevate the intimate cooperation between spacetime
geometry and matters to a quantum world has insinuated doubt into us
for the harmonious conspiracy. Therefore we will take a closer look
at (\ref{einstein-eq}) aiming to reveal an inmost conflict in
(\ref{einstein-eq}) behind the superficial harmony.

The first observation \cite{hsy-de} is that the gravitation
described by (\ref{einstein-eq}) presupposes a physically inviable
vacuum. If one consider a flat spacetime whose metric is given by
$g_{MN} = \eta_{MN}$, the left-hand side of (\ref{einstein-eq})
identically vanishes and so the right-hand side must cultivate a
completely empty space with $T_{MN} = 0$. But the concept of empty
space in Einstein gravity is in acute contrast to the concept of
vacuum in quantum field theory (QFT) where the vacuum is not empty
but full of quantum fluctuations. As a result, QFT claims that such
an empty space of nothing is inviable in Nature and, instead, the
vacuum is extremely heavy whose weight is maximally of Planck mass,
i.e., $\rho_{\mathrm{vac}}\sim M_P^4$. Thus, if there is no way to
completely suppress quantum fluctuations, Einstein gravity probably
either presupposes a physically inviable space or incorrectly
identifies the genetic origin of flat spacetime.

The second observation \cite{saharov} is that flat spacetime in
general relativity behaves like an elastic body with tension
although the flat spactime itself is the geometry of special
relativity. We have seen that slides for colloquium introducing
general relativity to the public or undergraduate students often
contain the image like Figure 1 below. The Figure 1 illustrates how
a massive body changes the geometry of spacetime around the mass
point. In general relativity, this (curved) geometry is described by
the gravitational field $g_{MN}(x) = \eta_{MN} + h_{MN}(x)$ and
interpreted as gravity. When the massive body moves to another
place, the original point where the body was placed will recover a
(nearly) flat geometry like a rubber band. That is, the (flat)
spacetime behaves like a metrical elasticity which opposes the
curving of space.
\begin{wrapfigure}{r}{0.4\textwidth}
  \vspace{-11pt}
  \begin{center}
    \includegraphics[width=0.4\textwidth]{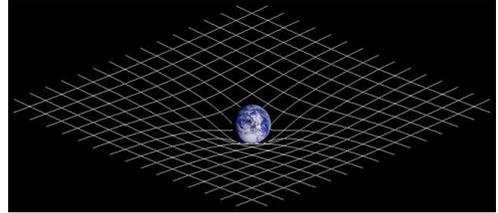}
  \end{center}
  \vspace{-11pt}
  \caption{\label{curvature}Two-dimensional analogy of spacetime
distortion. (Image from Wikipedia: Spacetime)}
  \vspace{-11pt}
\end{wrapfigure}
But this picture rather exhibits a puzzling nature of flat spacetime
because the flat spacetime should be a completely empty space
without any kind of energy as we remarked above. How is it possible
for an empty space of nothing to behave like an elastic body with
tension ? Moreover we know that the gravitational force is extremely
weak which implies that the space strongly withstands the curving
and so the tension of spacetime would be extremely big, maybe, of
the Planck energy.

The third observation \cite{zee} is that the gravitational field
$g_{MN}(x) = \eta_{MN} + h_{MN}(x)$ has a vacuum expectation value
(vev), i.e., $\langle g_{MN}(x) \rangle_{\mathrm{vac}} = \eta_{MN}$
like the Higgs field $\phi(x) = v + h(x)$. These two fields also
describe particles (either spin-2 graviton or spin-0 Higgs) as usual
quantum fields in Standard Model. However these two particles are
very exotic because all other fields, denoted as $\Psi$, in Standard
Model have a zero vev; $\langle \Psi \rangle_{\mathrm{vac}} = 0$. We
are reasonably understanding why the Higgs field has the nonzero vev
which triggers the electroweak symmetry breaking. In effect,
Standard Model is defined in a nontrivial vacuum with the Higgs
condensate $v = \langle \phi \rangle_{\mathrm{vac}}$ whose dynamical
scale is around $v \sim 246 \; {\rm GeV}$. Therefore any particle
interacting with the Higgs field feels a resistance in vacuum and
acquires a mass. What about the flat spacetime $\eta_{MN} = \langle
g_{MN}(x) \rangle_{\mathrm{vac}}$ ? Is it also originated from some
kind of vacuum condensate ? If so, what is the dynamical scale of
the condensate ? Note that the gravitation is characterized by its
own intrinsic scale given by the Newton constant $G = L_P^2$ where
$L_P = M_P^{-1} \sim 10^{-33}{\rm cm}$ is the Planck length and
classical gravity corresponds to $L_P \to 0$ limit.

The fourth observation \cite{ly-review,ly-jkps} is that gravity may
be not a fundamental force but an emergent force. A well-known fact
is that graviton is to a messenger particle of the gravitational
force as photon is to a messenger particle of the electromagnetic
force. Hence the graviton will not be a fundamental particle either,
if gravity is not a fundamental force. In general relativity the
gravitational force is represented by a Riemannian metric of curved
spacetime manifold $M$
\begin{equation} \label{metric}
\Big(\frac{\partial}{\partial s}\Big)^2 = g^{MN}(x)
\frac{\partial}{\partial x^M}
 \otimes \frac{\partial}{\partial x^N}.
\end{equation}
It is well-known that the metric (\ref{metric}) in the tetrad
formalism can be defined by the tensor product of two vector fields
$E_A = E_A^M (x) \frac{\partial}{\partial x^M} \in \Gamma(TM)$ as
follows
\begin{equation} \label{cartan}
\Big(\frac{\partial}{\partial s}\Big)^2  = \eta^{AB} E_A \otimes
E_B.
\end{equation}
Mathematically, a vector field $X$ on a smooth manifold $M$ is a
derivation of the algebra $C^\infty (M)$. Here the vector fields
$E_A \in \Gamma(TM)$ are the smooth sections of tangent bundle $TM
\to M$ which are dual to the vector space $E^A = E^A_M (x) dx^M \in
\Gamma(T^* M)$, i.e., $\langle E^A, E_B \rangle = \delta^A_B$. The
expression (\ref{cartan}) glimpses the avatar of gravity that a
spin-two graviton might arise as a composite of two spin-one {\it
vector fields}. In other words, the tensor product (\ref{cartan})
can be abstracted by the relation $(1 \otimes 1)_S = 2 \oplus 0$.
Note that any field $\Psi$ for fundamental particles in Standard
Model cannot be written as the tensor product of other two fields,
so to say, $\Psi = \Psi_1 \otimes \Psi_2$. Only composite particles
(or bound states) such as mesons can be represented in such a way.
Therefore graviton represented by the tensor product (\ref{cartan})
is certainly different from fundamental particles in Standard Model.

The final observation \cite{pad-1,pad-2} is that there is an acute
mismatch of symmetry between gravity and matters because gravity is
the only interaction sensitive to a shift of the Lagrangian by an
additive constant. To be precise, if one shift a matter Lagrangian
$\mathcal{L}_M$ by a constant $\Lambda$, that is,
\begin{equation}\label{shift-l}
\mathcal{L}_M  \to \mathcal{L}'_M  = \mathcal{L}_M - 2 \Lambda,
\end{equation}
it results in the shift of the energy-momentum tensor by $T_{MN} \to
T_{MN} - \Lambda g_{MN}$ in the Einstein equation
(\ref{einstein-eq}) although the equations of motion for matters are
invariant under the shift (\ref{shift-l}). Definitely the
$\Lambda$-term in (\ref{shift-l}) will appear as the cosmological
constant in Einstein gravity and it affects the spacetime structure.
For instance, a flat spacetime is no longer a solution of
(\ref{einstein-eq}). Even worse is that this clash of symmetry
brings about the stability problem of spacetime. As we remarked in
the first observation, the vacuum in QFT is a stormy sea of quantum
fluctuations which accommodates the vacuum energy of the order of
$M_P^4$. Fortunately the vacuum energy due to the quantum
fluctuations, regardless of how large it is, does not make any
trouble to QFT thanks to the symmetry (\ref{shift-l}). However the
general covariance requires that gravity couples universally to all
kinds of energy. Therefore the vacuum energy $\rho_{\mathrm{vac}}
\sim M_P^4$ will induce a highly curved spacetime whose curvature
scale $R$ would be $\sim M_P^2$ according to (\ref{einstein-eq}). If
so, the QFT framework in the background of quantum fluctuations must
be broken down due to a large back-reaction of background spacetime.
But we know that it is not the case. QFT is well-defined as ever in
the presence of the vacuum energy because the background spacetime
still remains flat, as we empirically know. What is wrong with this
argument ?

After consolidating all the suspicions inferred above, we throw a
doubt on the genesis that flat spacetime is free gratis, i.e., costs
no energy. All the above reasonings imply that the negligence about
the dynamical origin of flat spacetime defining a local inertial
frame in general relativity might be a core root of the
incompatibility inherent in (1). It should be remarked that the
genesis about spacetime cannot be addressed within the context of
general relativity because flat spacetime is a geometry of special
relativity rather than general relativity and so it is assumed to be
{\it a priori} given without reference to its dynamical origin. All
in all, it is tempted to infer that flat spacetime may be not free
gratis but a result of Planck energy condensation in vacuum
\cite{hsy-cc,hsy-ccc}. Surprisingly, if that inference is true, it appears as the
H\'oly Gr\'ail to cure several notorious problems in theoretical
physics; for example, to resolve the cosmological constant problem,
to understand the nature of dark energy and to explain why gravity
is so weak compared to other forces. After all, the target is to
formulate a background independent theory\footnote{Here we refer to
a background independent theory where any spacetime structure is not
{\it a priori} assumed but defined by the theory.} to correctly
explain the dynamical origin of flat spacetime. Note that Einstein
gravity is not completely background independent since it assumes
the prior existence of a spacetime manifold. But it turns out
\cite{hsy-ijmp-2009,hsy-jhep,ly-review} that the emergent gravity from
noncommutative (NC) geometry precisely realizes the desired
property, as will be surveyed in the next sections.

\section{Einstein gravity from electromagnetism on symplectic space}

Now we will show that the vierbeins in (\ref{cartan}) and so the
Riemannian metrics arise from electromagnetic fields living in a
space $(M, B)$ supporting a symplectic structure $B$
\cite{egnc-1,egnc-2,egnc-3,egnc-4}.\footnote{From now on, we will work in Euclidean space though
we still use the term ``spacetime". After illuminating how a space
is emergent from $U(1)$ gauge fields, we will speculatively touch
the issue of emergent time.} See
\cite{szabo-r,eg-review1,eg-review2,ly-review,eg-review3} for recent reviews
of this subject. The symplectic structure $B$ is a nondegenerate,
closed 2-form, i.e. $dB=0$
\cite{mechanics}. Therefore the symplectic structure $B$
defines a bundle isomorphism $B: TM \to T^*M$ by $X \mapsto A =
\iota_X B$ where $\iota_X$ is an interior product with respect to a
vector field $X \in \Gamma(TM)$. One can invert this map to obtain
the inverse map $\theta \equiv B^{-1}: T^*M \to TM$ defined by
$\alpha \mapsto X = \theta(\alpha)$ such that $X(\beta) =
\theta(\alpha,\beta)$ for $\alpha, \beta \in \Gamma(T^*M)$. The
bivector $\theta \in \Gamma(\Lambda^2 TM)$ is called a Poisson
structure of $M$ which defines a bilinear operation on
$C^\infty(M)$, the so-called Poisson bracket, defined by
\begin{equation}\label{poisson}
\{f, g\}_\theta = \theta(df, dg)
\end{equation}
for $f, g \in C^\infty(M)$. Then the real vector space
$C^\infty(M)$, together with the Poisson bracket $\{-, -\}_\theta$,
forms an infinite-dimensional Lie algebra, called a Poisson algebra
$\mathcal{P} =(C^\infty(M), \{-, -\}_\theta)$. First note that the
orthonormal tangent vectors $E_A = E^M_A(x)\partial_M \in
\Gamma(TM)$ satisfy the Lie algebra
\begin{equation}\label{lie-alg} \label{lie-vector}
[E_A, E_B] = - {f_{AB}}^C E_C.
\end{equation}
In general, the composition $[X,Y]$, the Lie bracket of $X$ and $Y$,
on $\Gamma(TM)$, together with the real vector space structure of
$\Gamma(TM)$, forms a Lie algebra $\mathcal{V} = (\Gamma(TM),
[-,-])$. There is a natural Lie algebra homomorphism between the Lie
algebra $\mathcal{V} = (\Gamma(TM), [-,-])$ and the Poisson algebra
$\mathcal{P} =(C^\infty(M), \{-, -\}_\theta)$ defined by
\cite{mechanics}
\begin{equation}\label{lie-homo}
  C^\infty(M) \to  \Gamma(TM) : f \mapsto X_f
\end{equation}
such that
\begin{equation}\label{ham-vec}
    X_f (g) = \theta(df, dg) = \{f, g\}_\theta
\end{equation}
for $f, g \in C^\infty(M)$. It is easy to prove the Lie algebra
homomorphism
\begin{equation} \label{vec-homo}
X_{\{f, g\}_\theta} = [X_f, X_g]
\end{equation}
using the Jacobi identity of the Poisson algebra $\mathcal{P}$.

Let us take $M = \mathbf{R}^4$ and a constant symplectic structure
$B = \frac{1}{2} B_{MN} dx^M \wedge dx^N$, for simplicity. A
remarkable point is that the electromagnetism on a symplectic
manifold $(\mathbf{R}^4, B)$ is completely described by the Poisson
algebra $\mathcal{P} =(C^\infty(M), \{-, -\}_\theta)$
\cite{hsy-epl,hsy-epj,hsy-jhep}. For example, the action is given by
\begin{equation}\label{nc-action}
    S = \frac{1}{4 g_{YM}^2} \int d^4 x \{ C_A, C_B \}_\theta^2
\end{equation}
where
\begin{equation}\label{coordinate-d}
    C_A(x) = B_{AB} x^B + A_A(x) \in C^\infty(M), \qquad A = 1, \cdots, 4
\end{equation}
are covariant dynamical coordinates describing fluctuations from the
Darboux coordinate $x^A$, i.e. $\{x^A, x^B \}_\theta = \theta^{AB}$,
and
\begin{eqnarray}\label{cc-field}
    \{ C_A(x), C_B(x) \}_\theta &=& -B_{AB} + \partial_A  A_B -
    \partial_B A_A + \{ A_A, A_B \}_\theta
    \nonumber \\
    &\equiv&  -B_{AB} + F_{AB}(x) \in C^\infty(M).
\end{eqnarray}
It is clear \cite{ly-review} that the equations of motion as well as
the Bianchi identity can be represented only with the Poisson
bracket $\{-, -\}_\theta$:
\begin{eqnarray} \label{eom-poisson}
&& \{ C^B(x), \{ C_A(x), C_B(x) \}_\theta \}_\theta  = 0, \\
\label{bi-poisson}
&& \{ C_A(x), \{ C_B(x), C_C(x) \}_\theta \}_\theta +
\mathrm{cyclic} = 0,
\end{eqnarray}
where
\begin{equation}\label{cov-f}
\{ C_A, \{ C_B, C_C \}_\theta \}_\theta = \partial_A F_{BC}
+ \{ A_A, F_{BC} \}_\theta \equiv D_A F_{BC} \in C^\infty(M).
\end{equation}

A peculiar thing for the action (\ref{nc-action}) is that the field
strength $F_{AB}$ in (\ref{cc-field}) is nonlinear due to the
Poisson bracket term although it is the curvature tensor of $U(1)$
gauge fields. Thus one can consider a nontrivial solution of the
following self-duality equation
\begin{equation}\label{u1-instanton}
    F_{AB} = \pm \frac{1}{2} {\varepsilon_{AB}}^{CD} F_{CD}.
\end{equation}
In fact, after the canonical Dirac quantization (\ref{q-homo}) of
the Poisson algebra $\mathcal{P} = (C^\infty(M), \{-, -\}_\theta)$,
the solution of the self-duality equation (\ref{u1-instanton}) is
known as noncommutative $U(1)$ instantons
\cite{nc-inst,hsy-nci1,hsy-nci2}. When applying the Lie algebra
homomorphism (\ref{vec-homo}) to (\ref{cc-field}), we get the
identity
\begin{equation}\label{field-vector}
    X_{F_{AB}} =  [V_A, V_B]
\end{equation}
where the vector fields $V_A
\equiv X_{C_A} \in \Gamma(TM)$ are obtained by the map (\ref{ham-vec})
from the set of the covariant coordinates $C_A(x)$ in
(\ref{coordinate-d}). As a result, the self-duality equation
(\ref{u1-instanton}) is mapped to the self-duality equation of the
vector fields $V_A$ \cite{hsy-epl,hsy-epj}:
\begin{equation}\label{u1-graviton}
    [V_A, V_B] = \pm \frac{1}{2} {\varepsilon_{AB}}^{CD} [V_C, V_D].
\end{equation}
Note that the vector fields $V_A = V_A^M \partial_M$ are divergence
free, i.e., $\partial_M V^M_A =0$ by the definition (\ref{ham-vec})
and so preserves a volume form $\nu$ because $\mathcal{L}_{V_A} \nu
= (\nabla \cdot V_A) \nu = 0$ where $\mathcal{L}_{V_A}$ is a Lie
derivative with respect to the vector field $V_A$. Furthermore it
can be shown \cite{hsy-jhep} that $V_A$ can be related to the
vierbeins $E_A$ by $V_A = \lambda E_A$ with $\lambda \in
C^\infty(M)$ to be determined.

If the volume form $\nu$ is given by
\begin{equation}\label{volume}
    \nu \equiv \lambda^{-2} \nu_g = \lambda^{-2} E^1 \wedge \cdots \wedge E^4
\end{equation}
or, in other words, $\lambda^2 = \nu(V_1, \cdots, V_4)$, one can
easily check that the triple of K\"ahler forms for a hyper-K\"ahler
manifold $M$ is given by \cite{ly-review}
\begin{equation}\label{ansatz}
    J_+^a = \frac{1}{2} \eta^a_{AB} \iota_A \iota_B \nu, \qquad
     J_-^{\dot{a}} = - \frac{1}{2} \overline{\eta}^{\dot{a}}_{AB} \iota_A \iota_B \nu,
\end{equation}
where $\iota_A$ is the interior product with respect to $V_A$ and
$\eta^a_{AB}$ and $\overline{\eta}^{\dot{a}}_{AB}$ are self-dual and
anti-self-dual 't Hooft symbols \cite{opy-jhep}. One can prove that
gravitational instantons satisfying the self-duality equation
\begin{equation}\label{g-self}
R_{MNAB} = \pm \frac{1}{2} {\varepsilon_{AB}}^{CD} R_{MNCD}
\end{equation}
are hyper-K\"ahler manifolds, i.e., $dJ_+^a = 0$ or $dJ_-^{\dot{a}}
= 0$ and vice versa. It is straightforward to prove \cite{ly-review}
that the hyper-K\"ahler conditions $dJ_+^a = 0$ or $dJ_-^{\dot{a}} =
0$ are precisely equivalent to (\ref{u1-graviton}) which can easily
be seen by applying to (\ref{ansatz}) the formula \cite{mechanics}
\begin{equation}\label{math}
    d(\iota_X \iota_Y \alpha) = \iota_{[X,Y]} \alpha + \iota_Y \mathcal{L}_X \alpha
    - \iota_X \mathcal{L}_Y \alpha + \iota_X \iota_Y d \alpha
\end{equation}
for vector fields $X, Y$ and a $p$-form $\alpha$.

In retrospect, the self-dual Lie algebra (\ref{u1-graviton}) was
derived from the self-duality equation (\ref{u1-instanton}) of
$U(1)$ gauge fields defined on the symplectic manifold
$(\mathbf{R}^4, B)$. As a consequence, $U(1)$ instantons on the
symplectic manifold $(\mathbf{R}^4, B)$ are gravitational instantons
\cite{sty-2006,ys-2006,hsy-epl,hsy-epj} ! We want to emphasize that the
emergence of Riemannian metrics from symplectic $U(1)$ gauge fields
is an inevitable consequence of the Lie algebra homomorphism between
the Poisson algebra $\mathcal{P} =(C^\infty(M), \{-, -\}_\theta)$
and the Lie algebra $\mathcal{V} = (\Gamma(TM), [-,-])$ if the
underlying action of $U(1)$ gauge fields is given by the form
(\ref{nc-action}). Moreover, the equivalence between $U(1)$
instantons in the action (\ref{nc-action}) and gravitational
instantons turns out to be a particular case of more general duality
between the $U(1)$ gauge theory on a symplectic manifold $(M,B)$ and
Einstein gravity \cite{hsy-jhep,hsy-siva}, as will be sketched
below.

First of all, we draw general results derived from the Lie algebra
isomorphism between the Poisson algebra $\mathcal{P} =(C^\infty(M),
\{-, -\}_\theta)$ and the Lie algebra $\mathcal{V} = (\Gamma(TM),
[-,-])$. Since $V_A = \lambda E_A \in
\Gamma(TM)$ where $\lambda^2 =
\det V_A^M$, the Riemannian metric (\ref{cartan}) is given by
\begin{equation} \label{em-imetric}
\Big(\frac{\partial}{\partial s}\Big)^2  = \delta^{AB} E_A \otimes
E_B = \lambda^{-2} \delta^{AB} V_A \otimes V_B
\end{equation}
or
\begin{equation} \label{em-metric}
ds^2  = \delta_{AB} E^A \otimes E^B = \lambda^{2} \delta_{AB} V^A
\otimes V^B
\end{equation}
where $V^A = \lambda^{-1} E^A \in \Gamma(T^*M)$ is a dual basis of
$V_A \in \Gamma(TM)$. Note that the smooth functions $C_A(x) \in
C^\infty(M) \; (A = 1, \cdots, 4)$ in (\ref{coordinate-d}) are
linearly independent and thus the vector fields $V_A \in \Gamma(TM)$
defined by (\ref{lie-homo}) are also linearly independent.
Accordingly the vector fields $V_A \; (A = 1, \cdots, 4)$ span a
full four-dimensional space. In effect, the metric (\ref{em-metric})
is completely determined by the set (\ref{coordinate-d}) of $U(1)$
gauge fields and it describes a general Riemannian manifold.

So far we did not impose the equations of motion (\ref{eom-poisson})
and the Jacobi identity (\ref{bi-poisson}) on the metric
(\ref{em-imetric}). Eventually we have to impose them because the
set (\ref{coordinate-d}) of $U(1)$ gauge fields obey
(\ref{eom-poisson}) and (\ref{bi-poisson}). In order to do that, let
us apply the Lie algebra homomorphism (\ref{vec-homo}) again to
(\ref{cov-f}) to yield
\begin{equation}\label{eom-geometry}
    X_{D_A F_{BC}} = [V_A, [V_B, V_C]] \in \Gamma(TM).
\end{equation}
It is then straightforward to get the following correspondence
\cite{hsy-jhep}
\begin{eqnarray} \label{corresp-eom}
D^B F_{AB} = 0 \qquad & \Leftrightarrow & \qquad
[V^B, [V_A, V_B]] = 0, \\
\label{corresp-bi}
D_A F_{BC} + \mathrm{cyclic} = 0 \qquad &
\Leftrightarrow & \qquad [V_A, [V_B, V_C]] + \mathrm{cyclic} = 0.
\end{eqnarray}
Now a critical question is whether the equations of motion
(\ref{corresp-eom}) for gauge fields together with the Jacobi (or
Bianchi) identity (\ref{corresp-bi}) can be written as the Einstein
equations for the metric (\ref{em-imetric}). A quick notice is that
(\ref{corresp-eom}) and (\ref{corresp-bi}) will end in some
equations related to Riemann curvature tensors because they
differentiate the metric (\ref{em-imetric}) twice.

To see what they are, recall that, in terms of covariant derivative,
the torsion $T$ and the curvature $R$ can be expressed as follows
\cite{big-book}
\begin{eqnarray} \label{torsion}
T(X, Y) &=& \nabla_X Y - \nabla_Y X - [X,Y], \\
\label{curvature}
R(X,Y)Z &=& [\nabla_X, \nabla_Y]Z - \nabla_{[X,Y]}Z,
\end{eqnarray}
where $X, Y$ and $Z$ are vector fields on $M$. Because $T$ and $R$
are multilinear differential operators, we get the following
relations \cite{gtp-naka}
\begin{eqnarray} \label{c-torsion}
T(V_A, V_B) &=& \lambda^2 T(E_A, E_B), \\
\label{c-curvature}
R(V_A, V_B) V_C &=& \lambda^3 R(E_A, E_B) E_C.
\end{eqnarray}
After imposing the torsion free condition $T(E_A, E_B) = 0$, it is
straightforward, using (\ref{c-torsion}) and (\ref{c-curvature}), to
derive the identity below
\begin{equation}\label{bi-id}
R(E_A, E_B) E_C + \mathrm{cyclic} = \lambda^{-3} \Big( [V_A, [V_B,
V_C]] + \mathrm{cyclic} \Big).
\end{equation}
Therefore we immediately see \cite{hsy-jhep} that the Bianchi
identity (\ref{corresp-bi}) for $U(1)$ gauge fields is equivalent to
the first Bianchi identity for Riemann curvature tensors, i.e.,
\begin{equation}\label{bi-bi}
D_A F_{BC} + \mathrm{cyclic} = 0 \qquad
\Leftrightarrow  \qquad R(E_A, E_B) E_C + \mathrm{cyclic} = 0.
\end{equation}

The mission for the equations of motion (\ref{corresp-eom}) is more
involved. An underlying idea is to carefully separate the right-hand
side of (\ref{corresp-eom}) into a part related to the Ricci tensor
$R_{AB}$ and a remaining part. Basically, we are expecting the
following form of the Einstein equations
\begin{equation}\label{einstein-teq}
D^B F_{AB} = 0 \qquad
\Leftrightarrow  \qquad    R_{AB} = 8\pi G \Big( T_{AB} - \frac{1}{2} \delta_{AB} T \Big).
\end{equation}
After a straightforward but tedious calculation \cite{hsy-jhep}, we
get a remarkably simple but cryptic result
\begin{equation} \label{emergent-einstein}
  R_{AB} = - \frac{1}{\lambda^2} \Big[ g^{(+)a}_D g^{(-)\dot{b}}_D
  \Big(\eta^a_{AC} \overline{\eta}^{\dot{b}}_{BC}
  + \eta^a_{BC} \overline{\eta}^{\dot{b}}_{AC} \Big) - g^{(+)a}_C g^{(-)\dot{b}}_D
  \Big(\eta^a_{AC} \overline{\eta}^{\dot{b}}_{BD}
  + \eta^a_{BC} \overline{\eta}^{\dot{b}}_{AD} \Big) \Big].
\end{equation}
To get the result (\ref{emergent-einstein}), we have defined the
structure equation of vector fields $V_A \in \Gamma(TM)$
\begin{equation}\label{streq-v}
    [V_A, V_B] = - {g_{AB}}^C V_C
\end{equation}
and the canonical decomposition
\begin{equation}\label{def-g}
g_{ABC} = g^{(+)a}_C \eta^a_{AB} + g^{(-)\dot{a}}_C
\overline{\eta}^{\dot{a}}_{AB}.
\end{equation}
A notable point is that the right-hand side of
(\ref{emergent-einstein}) consists of purely interaction terms
between self-dual and anti-self-dual parts in (\ref{def-g}) which is
the feature withheld by matter fields only \cite{oh-ya,loy}.
Incidentally, the self-duality equation (\ref{u1-graviton}) can be
understood as $g^{(-)\dot{a}}_C = 0$ (self-dual) or $g^{(+)a}_C = 0$
(anti-self-dual) in terms of (\ref{def-g}) and so $R_{AB} = 0$ in
(\ref{emergent-einstein}), i.e., (\ref{u1-graviton}) describes a
Ricci-flat manifold. Of course, this is consistent with the fact
that a gravitational instanton is a Ricci-flat, K\"ahler manifold.
Nevertheless a unique property of (\ref{emergent-einstein}) is to
contain a nontrivial trace contribution, i.e., a nonzero Ricci
scalar, due to the second part which is not existent in Einstein
gravity as was recently shown in \cite{loy}. By comparing
(\ref{emergent-einstein}) with (\ref{einstein-teq}), a surprising
content of the energy-momentum tensor was found in \cite{hsy-jhep},
which will be discussed in section 4.

We come to the conclusion that general relativity or gravity can
emerge from the electromagnetism supported on a symplectic spacetime
$(M, B)$, which is an interacting theory defined by the action
(\ref{nc-action}). How is it possible to realize the equivalence
principle or general covariance, the most important property in the
theory of gravity (general relativity), from the $U(1)$ gauge theory
on a symplectic or Poisson manifold ? It turns out
\cite{hsy-ijmp-2009,hsy-jhep} that the Poisson structure
(\ref{poisson}) of spacetime admits a novel form of the equivalence
principle even for the electromagnetic force, known as the Darboux
theorem or the Moser lemma in symplectic geometry, and consequently
the electromagnetism on a symplectic spacetime can be realized as a
geometrical property of spacetime. In the end, the symplectization
of spacetime geometry would be a novel and authentic way to quantize
gravity \cite{ly-review,eg-review3}.

\section{Noncommutative geometry and quantum gravity}

We have observed that Einstein gravity can be emergent from
electromagnetism as long as spacetime admits a symplectic structure
and its underlying theory is completely defined by the Poisson
algebra $\mathcal{P} = (C^\infty(M), \{-,-\}_\theta )$. For
instance, an underlying dynamical system for gravity is described by
the action (\ref{nc-action}) which leads to the equations of motion
(\ref{eom-poisson}). Note that the Jacobi identity
(\ref{bi-poisson}) is an important property for $\mathcal{P}$ to be
a Poisson algebra and to form a Lie algebra. One can understand the
Lie algebra morphism (\ref{lie-homo}) as the adjoint map defined by
\begin{equation}\label{adjoint-map}
    \mathrm{ad}_f : g \mapsto \{f, g \}_\theta
\end{equation}
and thence the action of any element on the algebra is a derivation,
i.e.,
\begin{equation}\label{derivation}
\mathrm{ad}_{f} (g \cdot h) = (\mathrm{ad}_f g) \cdot h + g \cdot \mathrm{ad}_f h
\end{equation}
for $f, g, h \in C^\infty(M)$. The Jacobi identity of the Poisson
algebra $\mathcal{P}$ is then equivalent to the identity
(\ref{vec-homo}) between the operators of the adjoint
representation. This identity implies that the map
(\ref{adjoint-map}) sending each element to its adjoint action is a
Lie algebra homomorphism of the original algebra $\mathcal{P}$ into
the Lie algebra $\mathcal{V}= (\Gamma(TM), [-,-])$ of its
derivations. This is a mathematical basis to explain how gravity is
emergent from the electromagnetism on a symplectic manifold $(M,
B)$.

Using the isomorphism between the Lie algebras $\mathcal{P}$ and
$\mathcal{V}$, we showed that a standard (commutative) dynamical
system for gravity can be described in terms of vector fields in
$\mathcal{V}$. Recall that vector fields on a usual (commutative)
space are derivations of the algebra $C^\infty(M)$ of smooth
functions on this space. And vector fields, being a global concept,
has its noncommutative (NC) generalization, called a derivation of
NC algebra. Now we will show how the derivation of NC algebra can be
obtained by canonically (\`a la Dirac) quantizing the Poisson
algebra $\mathcal{P} = (C^\infty(M), \{-,-\}_\theta )$.

The Dirac quantization of the Poisson algebra $\mathcal{P} =
(C^\infty(M), \{-,-\}_\theta )$ consists of a complex Hilbert space
$\mathcal{H}$ and of a quantization map $\mathcal{Q}$ to attach to
functions $f \in C^\infty(M)$ on $M$ operators $\widehat{f} \in
\mathcal{A}_\theta$ acting on $\mathcal{H}$ \cite{ncft1,ncft2}.
The map $\mathcal{Q}: C^\infty(M) \to \mathcal{A}_\theta$ by $f
\mapsto \mathcal{Q}(f) \equiv \widehat{f}$ should be $\mathbf{C}$-linear
and an algebra homomorphism:
\begin{equation}\label{q-homo}
    f \cdot g \mapsto \widehat{f \star g} = \widehat{f} \cdot \widehat{g}
\end{equation}
and
\begin{equation}\label{nc-homo}
    f \star g \equiv \mathcal{Q}^{-1} \Big(\mathcal{Q}(f) \cdot \mathcal{Q}(g) \Big)
\end{equation}
for $f, g \in C^\infty(M)$ and $\widehat{f}, \widehat{g}
\in \mathcal{A}_\theta$. The Poisson structure (\ref{poisson})
controls the failure of commutativity
\begin{equation}\label{q-comm}
    [ \widehat{f},  \widehat{g}] \sim i \{ f, g \}_\theta +
    \mathcal{O}(\theta^2).
\end{equation}
For example, the coordinate generators of $\mathcal{A}_\theta$ are
noncommuting with the Heisenberg algebra relation
\begin{equation}\label{heisenberg}
    [x^A, x^B] = i \theta^{AB}
\end{equation}
where we omit the hat for the coordinate generators for a notational
convenience. From the deformation quantization point of view, the NC
algebra of operators in $\mathcal{A}_\theta$ is equivalent to the
deformed algebra of functions defined by the Moyal $\star$-product
(\ref{nc-homo}) which is, according to the Weyl-Moyal map
\cite{ncft1,ncft2}, given by
\begin{equation}\label{star-prod}
    \widehat{f} \cdot \widehat{g} \cong (f \star g) (x) = \exp \Big(
    \frac{i}{2} \theta^{AB} \partial_A^x \partial_B^y \Big) f(x)
    g(y) |_{x=y}.
\end{equation}

According to the quantization map (\ref{q-homo}), every expressions
in $\mathcal{P}$ are mapped to corresponding operators in
$\mathcal{A}_\theta$. For instance, for the symplectic gauge fields
in (\ref{coordinate-d}), we have the map
\begin{equation}\label{nc-field}
C_A(x) \in C^\infty(M) \quad \Rightarrow \quad \widehat{C}_A(x) =
B_{AB} x^B + \widehat{A}_A(x) \in \mathcal{A}_\theta
\end{equation}
and, for the Poisson bracket in (\ref{cc-field}),
\begin{eqnarray}\label{q-map}
    \{ C_A(x), C_B(x) \}_\theta \; \Rightarrow \;
    -i [ \widehat{C}_A(x), \widehat{C}_B(x) ]_\star &=& - B_{AB}
    + \partial_A  \widehat{A}_B -
    \partial_B \widehat{A}_A - i [\widehat{A}_A,
    \widehat{A}_B]_\star \nonumber \\
    &\equiv&  -B_{AB} + \widehat{F}_{AB}(x) \in \mathcal{A}_\theta
\end{eqnarray}
where $[\widehat{f}, \widehat{g}]_\star =
\widehat{f} \star \widehat{g} - \widehat{g} \star \widehat{f}$.
The quantized action for NC $U(1)$ gauge fields is then given by
\begin{equation}\label{q-action}
    \widehat{S} = - \frac{1}{4 g_{YM}^2} \int d^4 x
    [\widehat{C}_A, \widehat{C}_B]_\star^2.
\end{equation}
Similarly, one can lift the adjoint map (\ref{lie-homo}) or
(\ref{adjoint-map}) to derivations of the NC algebra
$\mathcal{A}_\theta$:
\begin{equation}\label{nc-deriv}
    \mathrm{ad}^\star_{\widehat{f}} : \widehat{g} \mapsto -
    i[\widehat{f}, \widehat{g}]_\star
\end{equation}
that satisfies the Leibniz rule, i.e.,
\begin{equation}\label{leibniz}
    \mathrm{ad}^\star_{\widehat{f}} \; (\widehat{g}  \star \widehat{h})
    = (\mathrm{ad}^\star_{\widehat{f}} \; \widehat{g}) \star \widehat{h}
    + \widehat{g} \star \mathrm{ad}^\star_{\widehat{f}} \; \widehat{h}
\end{equation}
for $\widehat{f}, \widehat{g}, \widehat{h} \in \mathcal{A}_\theta$.
And the Jacobi identity of the $\star$-commutator (\ref{nc-deriv})
leads to the conclusion that the polydifferential operator on
$\mathcal{A}_\theta$ \cite{behr-sykora}, whose set is denoted as
$\Gamma_\theta(\widehat{TM})$,
\begin{equation}\label{poly-diff}
\mathrm{ad}^\star_{\widehat{f}} \equiv X^\star_{\widehat{f}} = X_f +
\sum_{n=2}^\infty \xi_{X_f}^{A_1 \cdots A_n} \partial_{A_1} \cdots
\partial_{A_n}
\end{equation}
is again a derivation of $\mathcal{A}_\theta$ satisfying the
deformed Lie algebra
\begin{equation}\label{d-lie-alg}
[X^\star_{\widehat{f}} \, , X^\star_{\widehat{g}}] =
X^\star_{[\widehat{f},
\, \widehat{g}]_\star}.
\end{equation}
It should be noted that the polydifferential operator
(\ref{poly-diff}) recovers the usual vector field in the commutative
limit $\theta \to 0$. Hence it is obvious that the left-hand side of
(\ref{d-lie-alg}) is a deformation of the ordinary Lie bracket of
vector fields. See \cite{behr-sykora} for an explicit formula up to
second order.

It is easy to ``quantize" the Lie algebra homomorphism
(\ref{eom-geometry}) using the above relation (\ref{d-lie-alg})
\begin{equation} \label{q-geometry}
    X^\star_{\widehat{D}_A \widehat{F}_{BC}}
    = [V^\star_A, [V^\star_B, V^\star_C]] \in \Gamma_\theta(\widehat{TM})
\end{equation}
where $V^\star_A \equiv X^\star_{\widehat{C}_A} \in
\Gamma_\theta(\widehat{TM})$ are generalized vector fields defined by (\ref{poly-diff}).
Accordingly we have a NC generalization of the correspondence
(\ref{corresp-eom}) given by \cite{hsy-jhep,ly-review}
\begin{eqnarray} \label{q-corresp-eom}
\widehat{D}^B \widehat{F}_{AB} = 0 \qquad & \Leftrightarrow & \qquad
[V^\star_B, [V^\star_A, V^\star_B]] = 0, \\
\label{q-corresp-bi}
\widehat{D}_A \widehat{F}_{BC} + \mathrm{cyclic} = 0 \qquad &
\Leftrightarrow & \qquad [V^\star_A, [V^\star_B, V^\star_C]] + \mathrm{cyclic} = 0.
\end{eqnarray}
Since the leading order in (\ref{poly-diff}) recovers the usual
vector fields, the Einstein equations (\ref{einstein-teq}) will
appear as the leading order of NC gauge fields described by
(\ref{q-corresp-eom}) and (\ref{q-corresp-bi}) and higher order
terms will generate derivative corrections of the Einstein gravity.
Although there is no concrete verification so far, it was
conjectured in \cite{hsy-ijmp-2009} that the resulting emergent
gravity from NC gauge fields will be based on the NC geometry
defined by
\begin{eqnarray} \label{nc-torsion}
\widehat{T}(X^\star, Y^\star) &=& \widehat{\nabla}_{X^\star} Y^\star
- \widehat{\nabla}_{Y^\star} X^\star - [X^\star, Y^\star], \\
\label{nc-curvature}
\widehat{R}(X^\star, Y^\star)Z^\star &=& [\widehat{\nabla}_{X^\star},
\widehat{\nabla}_{Y^\star}]Z^\star
- \widehat{\nabla}_{[X^\star, Y^\star]}Z^\star,
\end{eqnarray}
where $X^\star, Y^\star, Z^\star \in \Gamma_\theta(\widehat{TM})$
and $\widehat{\nabla}_{X^\star}$ is a generalized affine connection
on $\mathcal{A}_\theta$ evaluated at the vector field $X^\star$. If
so, the NC gravity in \cite{ncg1,ncg2} would be defined by NC gauge
fields.

Now we will argue that the NC geometry described by
(\ref{q-corresp-eom}) and (\ref{q-corresp-bi}) has to define a
quantum gravity at a microscopic scale, e.g., Planck scale $L_P$ and
provides a clue to realize a background independent formulation of
quantum gravity \cite{eg-review3}. One can meaningfully speak of a
NC dynamics (without reference to local concepts such as that of
points or time instant) provided that one describes NC dynamics in
terms of derivations of the corresponding algebra. This is precisely
the geometric basis underpinning the above construction. In this
approach, the crux for the background independentness is that the NC
spacetime defined by the Heisenberg algebra (\ref{heisenberg})
admits a separable Hilbert space $\mathcal{H}$ which is an
infinite-dimensional Fock space of two-dimensional quantum harmonic
oscillators. Therefore any NC fields in $\mathcal{A}_\theta$, which
are operators acting on $\mathcal{H}$, can be represented in the
Fock space $\mathcal{H}$ as $N \times N$ matrices where $N =
\mathrm{dim} \mathcal{H} \to \infty$ \cite{hsy-epj}. In the end, we have a matrix
representation, denoted as $\mathcal{A}_N$, for the NC $U(1)$ gauge
theory described by (\ref{q-corresp-eom}) and (\ref{q-corresp-bi})
that is given by \cite{nc-ikkt}
\begin{eqnarray} \label{m-corresp-eom}
\widehat{D}^B \widehat{F}_{AB} = 0
\qquad & \Leftrightarrow & \qquad
[\mathbf{C}^B, [\mathbf{C}_A, \mathbf{C}_B]] = 0, \\
\label{m-corresp-bi}
\widehat{D}_A \widehat{F}_{BC} + \mathrm{cyclic} = 0
\qquad & \Leftrightarrow & \qquad [\mathbf{C}_A, [\mathbf{C}_B, \mathbf{C}_C]]
+ \mathrm{cyclic} = 0,
\end{eqnarray}
where $\mathbf{C}_A = B_{AB} x^B + \mathbf{A}_A \in \mathcal{A}_N$
is a matrix representation in $\mathcal{H}$ of the NC field
$\widehat{C}_A(x) \in \mathcal{A}_\theta$. Note that the matrix
equations of motion (\ref{m-corresp-eom}) can be derived from the
0-dimensional IKKT matrix model \cite{ikkt,ikkt2} whose action is
given by
\begin{equation}\label{ikkt}
    S_M = - \frac{1}{4 g^2} \mathrm{Tr} [\mathbf{C}_A, \mathbf{C}_B]^2.
\end{equation}

An underlying picture for the emergent gravity \cite{ly-review} will
become clear by recasting the arguments so far with the matrix
action (\ref{ikkt}). The action (\ref{ikkt}) is zero-dimensional and
so it does not assume any kind of spacetime structure. There are
only four $N \times N$ Hermitian matrices $\mathbf{C}_A \; (A =1,
\cdots, 4)$ which are subject to a couple of algebraic relations
defined by the right-hand sides of (\ref{m-corresp-eom}) and
(\ref{m-corresp-bi}). Therefore, in order to {\it create a Universe
(or any existence)}, first we have to specify a vacuum of the theory
where all fluctuations are supported. For consistency, the vacuum
should also satisfy (\ref{m-corresp-eom}) and (\ref{m-corresp-bi}).
Since the action (\ref{ikkt}) allows infinitely many solutions even
with different topologies, it is not unique but there is a natural
``primitive" vacuum defined by
\begin{equation}\label{vacuum}
    \langle \mathbf{C}_A \rangle_{\mathrm{vac}} \equiv
    \widehat{A}^{(0)}_A = B_{AB} x^B
\end{equation}
where $B_{AB}$ is a constant matrix of rank 4. The vacuum
(\ref{vacuum}) describes a uniform condensate of NC gauge fields and
obeys (\ref{m-corresp-eom}) and (\ref{m-corresp-bi}) if $x^B
\in \mathcal{A}_N$ satisfy the Heisenberg algebra (\ref{heisenberg}).
We can also introduce fluctuations over the vacuum (\ref{vacuum})
which are represented by $\mathbf{A}_A$. Because there is a Hilbert
space $\mathcal{H}$ as a representation space of the Heisenberg
algebra (\ref{heisenberg}), we can regard the matrices $\mathbf{C}_A
\in \mathcal{A}_N$ as operators in $\mathcal{A}_\theta$ acting on
the Hilbert space $\mathcal{H}$. According to the Weyl-Moyal map
(\ref{star-prod}), these adjoint operators are in turn mapped to NC
fields represented by (\ref{nc-field}). It is worthy of remark that
the pith for the duality between geometry and algebra originates
from the coherent condensation (\ref{vacuum}) of quantum harmonic
oscillators in vacuum, which grants a symplectic structure to the
vacuum. It is well-known \cite{szabo-r} that the NC algebra
$\mathcal{A}_\theta$ generated by (\ref{heisenberg}) admits a
nontrivial inner automorphism whose infinitesimal form is called an
inner derivation defined by (\ref{nc-deriv}). As a result, the
dynamics of fluctuations on the vacuum (\ref{heisenberg}) can always
be described by the inner derivations of the algebra
$\mathcal{A}_\theta$, as was verified in (\ref{q-corresp-eom}) and
(\ref{q-corresp-bi}). We showed that their commutative limit is
nothing but the Einstein gravity or general relativity.

Now we are ready to disclose the secret nature of spacetime we have
posed in the introduction. Because the IKKT matrix model
(\ref{ikkt}) does not assume any prior existence of spacetime from
the beginning, in other words, (\ref{ikkt}) is a background
independent theory, it is necessary to define a configuration in the
algebra $\mathcal{A}_\theta$, for instance, like (\ref{vacuum}), to
generate any kind of spacetime structure, even for flat spacetime.
So the question is: What is the spacetime emergent from the vacuum
(\ref{vacuum}) which signifies a uniform condensate of NC gauge
fields in vacuum ? The definition (\ref{nc-deriv}) immediately says
that the corresponding vector field $V_A^{(0)} =
\delta^B_A \partial_B$ for the vacuum gauge field $\widehat{A}^{(0)}_A$
is precisely that of flat spacetime, i.e., $\langle g_{AB}
\rangle_{\mathrm{vac}}= \delta_{AB}$.
See (\ref{em-metric}) for the metric where $\lambda^2 =1$ in this
case. Remarkably the vacuum (\ref{vacuum}) responsible for the flat
spacetime is not an empty space unlike general relativity. Instead
the flat spacetime is emergent from a uniform condensation of gauge
fields in vacuum \cite{hsy-jhep,ly-review}. Its surprising
consequences will be discussed in next section.

\section{Emergent spacetime and dark energy}

To recapitulate, it was shown that the symplectic structure of
spacetime $M$ leads to an isomorphism between symplectic geometry
$(M, B)$ and Riemannian geometry $(M,g)$ where the deformations of
symplectic structure $B$ in terms of electromagnetic fields $F=dA$
are transformed into those of Riemannian metric $g$. This approach
for quantum gravity allows a background independent formulation
which provides a novel and authentic way to quantize gravity
\cite{eg-review3}. As such, the theory should be formulated in a way
that the spacetime geometry arises from a vacuum solution to the
field equation of the theory. One should not have to specify a
preferred background spacetime in order to write down the field
equations. The theory (\ref{ikkt}) described by large $N$ matrices
is precisely the case. Indeed, the flat spacetime arises from the
vacuum solution (\ref{vacuum}). All other fluctuations over the
vacuum are represented by NC $U(1)$ gauge fields with the action
(\ref{q-action}), whose commutative limit $\theta \to 0$ reduces to
the action (\ref{nc-action}) we will mostly focus on. Only these
fluctuations will deform the vacuum geometry as $
\partial_A \to E_A =  \partial_A + h_A^M(x)
\partial_M$ or $\delta_{AB} \to g_{AB} = \delta_{AB} + h_{AB}$,
according to the map (\ref{lie-homo}). One immediate conclusion is,
thus, that a uniform condensation of gauge fields such as the vacuum
(\ref{vacuum}) does {\it not} gravitate
\cite{hsy-cc,hsy-jhep} because it is simply responsible for the
generation of flat spacetime. A natural question is then what is the
dynamical scale of the vacuum condensate (\ref{vacuum}).\footnote{We
do not have an understanding yet how time emerges in this context.
Therefore it may be rude to talk about the dynamics, energy, etc.
But a qualitative nature of the underlying physics can equally be
addressed even in Euclidean space. Hence, for some time, we will
reluctantly stay in Euclidean space not to get astray due to the
notorious issue of emergent time.}

Because gravity emerges from NC $U(1)$ gauge fields described by the
action (\ref{nc-action}) or (\ref{q-action}), the parameters,
$g_{YM}$ and $\theta = \big( \frac{1}{B} \big)$, defining the NC
gauge theory should be related to the Newton constant $G$ in
emergent gravity. A simple dimensional analysis leads to the
relation \cite{hsy-jhep}
\begin{equation}\label{newton}
    \frac{G \hbar^2}{c^2} \sim g^2_{YM} |\theta|.
\end{equation}
Then one can immediately estimate the vacuum energy
$\rho_{\mathrm{vac}}$ caused by the condensate (\ref{vacuum}):
\begin{equation}\label{v-energy}
    \rho_{\mathrm{vac}} \sim \frac{1}{g^2_{YM}} |B_{AB}|^2 \sim
    g^2_{YM} M_P^4 \sim 10^{-2} M_P^4
\end{equation}
where $M_P = (8 \pi G)^{-1/2} \sim 10^{18} \mathrm{GeV}$ is the
Planck mass and $g^2_{YM} \sim \frac{1}{137}$. Finally the emergent
gravity reveals a remarkable picture that the condensation of Planck
energy in vacuum is actually the origin of flat spacetime. That is
to say, the huge Planck energy (\ref{v-energy}) was simply used to
make a flat spacetime. Hence we can conclude that the vacuum energy
$\rho_{\mathrm{vac}} \sim M_P^4$ does not gravitate, which is a
tangible difference from Einstein gravity. It is of prime importance
that the emergent gravity should not contain a coupling of
cosmological constant like $\int d^4 x \sqrt{g} \Lambda$. This
important conclusion may be more strengthened by looking at the
definition (\ref{lie-homo}) of emergent metric which is insensitive
to a constant vacuum energy because it requires the identity
\begin{equation}\label{vec-f-id}
    X_{F_{AB} - B_{AB}} = X_{F_{AB}} \in \Gamma(TM)
\end{equation}
for a constant field strength $B_{AB}$. Consequently, the emergent
gravity clearly dismisses the cosmological constant problem
\cite{hsy-cc,hsy-de}.

As a necessary consequence, the emergent gravity respects the shift
symmetry (\ref{shift-l}). For example, under a shift in the
$B$-field, $B \to B' = B + b$, with $b$ a constant antisymmetric
tensor, the NC field theory (\ref{q-action}) defined in the new
background $\theta' = \big(\frac{1}{B+b}
\big)$ is physically equivalent to that in the old one $\theta = \big(\frac{1}{B}
\big)$ due to the well-known Seiberg-Witten equivalence \cite{sw-nc}.
Moreover the Hilbert spaces $\mathcal{H}(\theta')$ and
$\mathcal{H}(\theta)$ for the Heisenberg algebra (\ref{heisenberg})
are isomorphic to each other. Also the vector fields $V'^{(0)}_A$
and $V^{(0)}_A$ determined by $B'$ and $B$ backgrounds are equally
flat as long as they are constant. This means that the shift
symmetry (\ref{shift-l}) belongs to a global automorphism (a Darboux
transformation) of the NC algebra $\mathcal{A}_\theta$, i.e.
$\mathcal{A}_{\theta'} \cong \mathcal{A}_\theta$, which can be
interpreted as a global Lorentz transformation \cite{ly-review}.

If a flat spacetime emerges from the Planck energy condensation
(\ref{v-energy}) in vacuum,\footnote{We may emphasize that it
 is an inevitable consequence if quantum gravity should be formulated
 in a background independent way and so the spacetime geometry emerges
 from a vacuum configuration of some fundamental ingredients in the theory.
 It is then reasonable to assume that the gravitational constant $G = M_P^{-2}$ would
set a natural dynamical scale for the emergence of gravity and
spacetime.} we can draw several interesting implications (though
necessarily speculative) which seem to resolve all the puzzles posed
in the introduction. First of all, it implies that spacetime will
behave like an elastic body with the tension of Planck energy
\cite{saharov}. In other words, gravitational fields generated by the
deformations of the background (\ref{vacuum}) will be very weak
because the spacetime vacuum is very solid with a stiffness of the
Planck mass. Therefore the dynamical origin of flat spacetime
explains the metrical elasticity opposing the curving of space as
depicted in Figure 1 and the stunning weakness of gravitational
force
\cite{hsy-de}. Furthermore the emergent spacetime implies that the
global Lorentz symmetry, being an isometry of flat spacetime, should
be a perfect symmetry up to the Planck scale because the flat
spacetime was originated from the condensation of the maximum energy
in Nature.

The emergence of spacetime by the vacuum condensate (\ref{vacuum})
probably also has very interesting implications to cosmology
\cite{ly-jkps}. It is worthwhile to notice that the vacuum algebra
(\ref{heisenberg}) describes an extremely coherent condensation
because it is the Heisenberg algebra of two-dimensional quantum
harmonic oscillator. As a result, the spacetime vacuum
(\ref{vacuum}) should describe a zero-entropy state in spite of the
involvement of Planck energy like as the electrical resistance of
superconductors is zero because the Cooper pair condensate moves as
a coherent quantum mechanical entity. It is very mysterious but it
should be the case, because the flat spacetime is a completely empty
space from the viewpoint of general relativity and so has no
entropy. The thermodynamic laws then suggest that a global time
evolution of universe will have an arrow since the entropy of
universe has to increase anyway since its birth. It was argued in
\cite{ly-jkps} that the coherent condensate (\ref{vacuum}) of
spacetime ``monads" may explain {\it the arrow of time} in the
cosmic evolution of our universe.

One may envisage a spontaneous creation of an exponentially
expanding universe. But one could not say that the de Sitter
universe was created out of field energy in a preexisting space. If
we intend to understand the Planck energy condensation in vacuum as
a dynamical process, the time scale for the condensate will be
roughly of the Planck time. Thus it is natural to consider that the
explosive inflation era that lasted roughly $\mathrm{10^{-33}}$
seconds at the beginning of our univese corresponds to the dynamical
process for the instantaneous condensation of vacuum energy
$\rho_{\mathrm{vac}} \sim M_P^4$ to enormously spread out a
spacetime. A simple consideration of energy conservation for the
condensate leads to the expansion rate
\begin{equation}\label{inflation}
    \frac{1}{2} V_I^2 - \frac{4 \pi G \rho_{\mathrm{vac}}}{3} R^2 =
    0  \qquad \Rightarrow \qquad V_I = H_I R
\end{equation}
where $H_I = \sqrt{\frac{8 \pi G \rho_{\mathrm{vac}}}{3}} \sim M_P$.
This implies that the dynamical process for the vacuum condensate
may explain a cosmic inflation to generate an extremely large
spacetime $\sim e^{60} L_H$. Unfortunately, it is not clear how to
microscopically describe this dynamical process by using the matrix
action (\ref{ikkt}).\footnote{Recently there was a very interesting
work \cite{knt-matrix,knt-matrix2} addressing this issue using the
Monte Carlo analysis of the type IIB matrix model in Lorentzian
signature. In this work it was found that three out of nine spatial
directions start to expand at some critical time after which exactly
3+1 dimensions dynamically become macroscopic.} Nevertheless, it is
quite obvious that the cosmic inflation should be a dynamical
condensation in vacuum for the generation of spacetime according to
our emergent gravity picture \cite{ly-jkps}.

Note that the vacuum condensate (\ref{vacuum}) causes the
microscopic spacetime to be NC and so introduces a spacetime
uncertainty relation. Therefore, a further accumulation of energy
over the NC spacetime (\ref{heisenberg}) will be subject to the
UV/IR mixing \cite{uvir}. This spacetime exclusion will prevent a
very localized energy from further condensing into the vacuum, which
may correspond to the stability of spacetime. This reasoning implies
that the condensation of the vacuum energy (\ref{v-energy}) happened
at most only once. By the same reason, the cosmic inflation should
take place only once and, thereby, eternal inflation and cyclic
universe seem to be inconsistent with our picture \cite{ly-jkps}.

The special relativity unifies space and time into a single entity
-- spacetime. Hence it will be desirable to put space and time on an
equal footing. If a space is emergent, so should time. But the
concept of time is more stringent since it is difficult to give up
the causality and unitarity. We believe that a naive introduction of
NC time, e.g., $[t, x] = i\theta$, will be problematic because it is
impossible to keep the locality in time with the NC time and so to
protect the causality and unitarity. How can we define the emergent
time together with the emergent space ? How is it entangled with the
space to unfold into a single entity -- spacetime and to take the
shape of Lorentz covariance ? A decisive lesson comes from quantum
mechanics.

Quantum mechanics is the formulation of mechanics in NC phase space
\begin{equation} \label{qm}
[x^i, p_j] = i \hbar \delta^i_j.
\end{equation}
In quantum mechanics, the time
evolution of a dynamical system is described by the quantum
Hamilton's (Heisenberg) equation
\begin{equation}\label{h-eq}
 i \hbar   \frac{d}{dt} \widehat{f}(t) = [\widehat{f}(t), \widehat{H}]
\end{equation}
where $\widehat{f}(t), \widehat{H} \in \mathcal{A}_\hbar$ are
operators in the set of observables $\mathcal{A}_\hbar$. This
equation can be integrated to give a finite time evolution given by
\begin{equation}\label{time-ev}
\widehat{f}(t) = U(t)^\dagger \widehat{f}(0) U(t)
\end{equation}
where $U(t) = \exp (-i \widehat{H} t /\hbar)$ is the time evolution
operator. Therefore, as we know well, an intrinsic (particle) time
in quantum mechanics is defined as an inner automorphism of NC
algebra $\mathcal{A}_\hbar$. Note that the time evolution
(\ref{time-ev}) is meaningful only if the underlying algebra
$\mathcal{A}_\hbar$ is NC but it does not require an operator or NC
time to describe the history of the particle system.

A notable point is that any Poisson manifold $(M, \theta)$ always
admits a dynamical Hamiltonian system on $M$ where the Poisson
structure $\theta$ is a bivector in $\Gamma(\Lambda^2 TM)$ and the
dynamics of a system is described by the Hamiltonian vector field
$X_f = \theta(df)$ of an underlying Poisson algebra
\cite{mechanics}. After a (Dirac or deformation) quantization of the
Poisson algebra, one can describe the dynamics of the system in
terms of derivations of an underlying NC algebra. For example, in
the classical limit $\hbar \to 0$, the Heisenberg equation
(\ref{h-eq}) reduces to the classical Hamilton's equation
\begin{equation}\label{c-hamilton}
 \frac{d}{dt} f(t) = X_H (f(t)) = \{ f(t), H \}_\hbar
\end{equation}
where $X_H$ is a Hamiltonian vector field defined by $\iota_{X_H}
\omega = dH$ with the symplectic structure $\omega = \sum_i dx^i
\wedge dp_i$ of the particle phase space. In order to get an
insight about the emergent time, it would be worthwhile to realize
that the mathematical structure of emergent gravity is basically the
same as quantum mechanics. The former is based on the NC space
(\ref{heisenberg}) while the latter is based on the NC phase space
(\ref{qm}).

Therefore we can also apply the same philosophy to the case of NC
spacetime \cite{hsy-jhep}. We suggest the concept of time in
emergent gravity as the corresponding Hamilton's equation in the NC
space (\ref{heisenberg})
\begin{equation}\label{nch-eq}
 \frac{d}{dt} \widehat{f}(t) = -i [\widehat{f}(t),
 \widehat{H}]_\star
\end{equation}
where $\widehat{f}(t), \widehat{H} \in \mathcal{A}_\theta$. Its
finite version will be given by an inner automorphism, similar to
(\ref{time-ev}), of the NC algebra $\mathcal{A}_\theta$. In the
commutative limit $\theta \to 0$, (\ref{nch-eq}) reduces to the
similar equation as (\ref{c-hamilton})
\begin{equation}\label{c-time}
 \frac{d}{dt} f(t) = X_H (f(t)) = \{ f(t), H \}_\theta
\end{equation}
where $X_H$ is a Hamiltonian vector field defined by $\iota_{X_H} B
= dH$. Conversely, the Heisenberg equation (\ref{nch-eq}) is the
quantization of the classical Hamilton's equation (\ref{c-time})
following the rule (\ref{derivation}). But, if the Hamiltonian
$\widehat{H}$ is time-independent, one can infer by analogy with
quantum mechanics that the Heisenberg equation (\ref{nch-eq})
describes only an internal time evolution over the space.

Let us consider a dynamical evolution described by the change of a
symplectic structure from $\omega_0 = B$ to $\omega_t = \omega_0 +
t(\omega_1 - \omega_0)$ for all $ 0 \leq t \leq 1$ where $\omega_1 -
\omega_0 = dA$. A remarkable point (due to the Moser lemma \cite{mechanics}) is that
there exists a one-parameter family of diffeomorphisms $\phi: M
\times \mathbf{R} \to M$ such that $\phi^*_t (\omega_t) = \omega_0,
\; 0 \leq t \leq 1$. Then the evolution of the symplectic structure
is locally described by the flow $\phi_t$ starting at $\phi_0 =
\mathrm{identity}$ and generated by time dependent vector fields
$X_t = \frac{d \phi_t}{dt} \circ \phi_t^{-1}$ satisfying the
equation
\begin{equation}\label{moser}
    \iota_{X_t} \omega_t + A = 0.
\end{equation}
Actually the covariant coordinates $X^A(x) = \theta^{AB} C_B (x) =
x^A + \theta^{AB} A_B (x)$ in (\ref{coordinate-d}) correspond to the
Darboux transformation $\phi_1: x^A \mapsto X^A$ obeying $\phi_1^*
(B+F) = B$ \cite{hsy-ijmp-2009}. Note that the emergence of gravity
originates from the global existence of the one-parameter family of
diffeomorphisms $\phi_t
\in \mathrm{Diff}(M)$ describing the local deformation of an initial
symplectic structure $\omega_0 = B$ due to the electromagnetic force
$F= dA$. Here we observe that the fluctuation of background geometry
(determined by $\omega_0 = B$) due to the deformation of symplectic
structures necessarily accompanies the time evolution of entire
geometry.

There are two facts known in symplectic geometry material to the
concept of emergent time. The time evolution of a time-dependent
system can again be defined by the inner automorphism of an extended
phase space whose extended Poisson bivector is given by
\cite{mechanics}
\begin{equation}\label{ex-poisson}
    \widetilde{\theta} = \theta + \frac{\partial}{\partial t} \wedge
    \frac{\partial}{\partial H}.
\end{equation}
As usual, the generalized Hamiltonian vector field is defined by
\begin{equation}\label{g-vec}
\widetilde{X}_H = - \widetilde{\theta}(dH) =
\theta^{AB} \frac{\partial H}{\partial x^B} \frac{\partial}{\partial
x^A} + \frac{\partial}{\partial t}
\end{equation}
and the corresponding Hamilton's equation is given by
\begin{equation} \label{g-ham}
    V_0 (f) \equiv \widetilde{X}_H (f)
    = \{ C_0, f \}_{\widetilde{\theta}} = \frac{\partial f}{\partial t}
    + \{ C_0, f \}_\theta
\end{equation}
where $C_0 \equiv - H \in C^\infty(M \times \mathbf{R})$. Basically
we regard $C_0 = - H$ as an another dynamical variable and so we
have a set of coordinates denoted by $\widetilde{C}_{\widetilde{A}}
= (C_0, C_A)$ in $M \times \mathbf{R}$ and, following the same
philosophy as (\ref{ham-vec}), we have introduced a vector field
$V_0 = \widetilde{X}_H \in \Gamma (T(M \times \mathbf{R}))$. But
note that the original vector fields $V_A \equiv X_{C_A}$ remain
intact because of the relation $\widetilde{V}_A =
\widetilde{\theta}(dC_A) = \theta(dC_A) = V_A$. Another important
point is the theorem of Souriau and Sternberg \cite{sternberg}
stating that a nontrivial time evolution in the presence of
electromagnetic fields can be described by the Hamilton's equation
with a free Hamiltonian $H = H_0$ but with a new Poisson structure
$\Theta = \Big(\frac{1}{B + F} \Big)$ deformed by the
electromagnetic force $F= dA$. In the case of (\ref{g-ham}), this
theorem means \cite{ly-review} that
\begin{equation}\label{ss-theorem}
 V_0 (f) = \frac{\partial f}{\partial t}
    - \{ H_0, f \}_\Theta
\end{equation}
where $\Theta^{AB} = \{X^A, X^B\}_\theta$ and $H_0$ is a Hamiltonian
function when $F=0$, viz., for flat spacetime.

The result (\ref{ss-theorem}) reveals a consistent picture with
general relativity about time \cite{hsy-jhep}. If $\Theta^{AB}$ is
constant (homogeneous), e.g. $\Theta^{AB}=\theta^{AB}$, a clock will
tick everywhere at the same rate because (\ref{ss-theorem}) is
exactly the same as the time evolution on flat spacetime. But, if
$\Theta^{AB}(x)$ is not constant (inhomogeneous) and so an
underlying geometry is curved, the time evolution will not be
uniform and a clock will tick at the different rate at different
places. Also it is quite plausible that the local Lorentz symmetry
would be recovered on a local chart because $\Theta^{AB}(x)$ on the
local chart will not be significantly changed and thus the time
evolution there is locally the same as the flat spacetime.

The above picture can be more illuminated by evaluating the metric
on $M \times \mathbf{R}$ defined by the vector fields $(V_0, V_A)
\in \Gamma (T(M \times \mathbf{R}))$ or their dual one-forms $(V^0, V^A) \in \Gamma
(T^*(M \times \mathbf{R}))$. The resulting (4+1)-dimensional metric is given by
\begin{equation}\label{4+1}
    ds^2 = \lambda^2 \Big( - dt^2 + V^A_M V^A_N (dx^M - \mathbf{A}^M)
    (dx^N - \mathbf{A}^N) \Big)
\end{equation}
where $\mathbf{A}^M = - \Big( \theta^{MN} \frac{\partial
C_0}{\partial x^N} \Big) dt$ and $\lambda^2 = \nu(V_0, V_1, \cdots,
V_4)$ with volume form $\nu = dt \wedge d^4 x$. One can easily see
that the metric (\ref{4+1}) reduces to the (4+1)-dimensional
Minkowski spacetime after turning off all fluctuations.\footnote{We
do not understand why the time emerges with an opposite sign, i.e.,
with the Minkowski signature. The signature is just our wishful
choice.} Interestingly, the metric (\ref{4+1}) appears as an
emergent geometry of matrix quantum mechanics -- the BFSS matrix
model \cite{bfss} -- whose action is given by
\begin{equation}\label{bfss}
    S_{MQM} = \frac{1}{g^2} \int dt \mathrm{Tr} \Big( -
    \frac{1}{2} \big( \mathbf{D}_0 \mathbf{C}_A \big)^2 +
    \frac{1}{4}[\mathbf{C}_A, \mathbf{C}_B]^2 \Big)
\end{equation}
where $\mathbf{D}_0 \mathbf{C}_A = \frac{\partial
\mathbf{C}_A}{\partial t} - i [ \mathbf{A}_0, \mathbf{C}_A]$.
Using the relationship between large $N$ matrix model and NC field
theory under the Moyal vacuum (\ref{vacuum}) with $\langle
\mathbf{A}_0 \rangle_{\mathrm{vac}} = 0$, one can show \cite{hsy-epj} that the
matrix quantum mechanics (\ref{bfss}) is equivalent to
(4+1)-dimensional NC $U(1)$ gauge theory. From the NC gauge theory
representation of the matrix model (\ref{bfss}), it is
straightforward to reproduce the emergent metric (\ref{4+1}) using
the vector fields $V_A \in \Gamma(TM)$ determined by NC fields. In
this way, we may get some deep insight about the formidable issue of
emergent time \cite{ly-review}.

Now let us return to the result (\ref{emergent-einstein}) to discuss
a content of the energy-momentum tensor defined by its right-hand
side. First it will be convenient to decompose the right-hand side
(\ref{emergent-einstein}) into two parts \cite{hsy-jhep}:
\begin{eqnarray} \label{em-maxwell}
  8 \pi G T_{AB}^{(M)} &=& - \frac{1}{\lambda^2} \Big( g_{ACD} g_{BCD}
  - \frac{1}{4} \delta_{AB} g_{CDE} g_{CDE} \Big), \\
  \label{em-de}
  8 \pi G T_{AB}^{(L)} &=& \frac{1}{2 \lambda^2} \Big( \rho_{A}
  \rho_{B} - \Psi_{A} \Psi_{B}
  - \frac{1}{2} \delta_{AB} \big( \rho^2_{C} - \Psi^2_{C} \big) \Big),
\end{eqnarray}
where $\rho_A \equiv g_{BAB}$ and $\Psi_A \equiv - \frac{1}{2}
\varepsilon^{ABCD} g_{BCD}$.
Recall that the result (\ref{emergent-einstein}) was obtained in
Euclidean space. In order to get a corresponding result in
(3+1)-dimensional Lorentzian spacetime just like above, we need to
start with a three-dimensional NC space. But we cannot complete a
full three-dimensional NC space with the Moyal algebra
(\ref{heisenberg}) since it is possible only with even dimensions.
Instead, it may be necessary to have a Lie algebra vacuum
\cite{hsy-siva}, e.g. $[x^A, x^B] = i \varepsilon^{ABC} x^C \;(A,B,C
= 1,2,3)$, or a Nambu vacuum, i.e. $[x^A, x^B, x^C] = i
\varepsilon^{ABC}$. Unfortunately, the calculation for these cases
is much more difficult. Even it is quite demanding to define
derivations (i.e., vector fields) for the latter case although the
former case is rather well-known from the representation theory of
Lie algebra. Therefore we will take a simple-minded recipe -- the
Wick rotation. (We are not happy with this trick.)

The Wick rotation will be defined by $x^4 = i x^0$. Under this Wick
rotation, we get the results $\delta_{AB} \to \eta_{AB}, \;
\varepsilon^{1234} = 1 \to -\varepsilon^{0123} = -1$ and $\Psi_A \to i \Psi_A$
in Minkowski spacetime. There are some reasons that the
energy-momentum tensor (\ref{em-maxwell}) has to be mapped to the
one of the usual Maxwell theory in commutative spacetime. Indeed it
was argued in \cite{hsy-jhep} that it can be done by reversing the
map (\ref{lie-homo}). But, as we already remarked in section 2, the
engrossing part is (\ref{em-de}) since it is absent in Einstein
gravity and would be a rather unique feature of emergent gravity.
Since we are eventually interested in a long-wavelength limit, we
will take only the scalar mode in (\ref{em-de}) which will be a
source of the expansion/contraction of spacetime. This will be a
good approximation since the remaining term corresponds to a
quadruple mode which gives rise to the shear distortion of spacetime
and can thus be neglected. For the same reason, we can ignore the
Maxwell energy-momentum tensor (\ref{em-maxwell}) since it is also a
purely quadruple mode. Hence, in the long-wavelength limit where
$\langle \rho_M \rho_N \rangle \simeq \frac{1}{4} g_{MN} \rho_P^2$
and $\langle \Psi_M \Psi_N \rangle \simeq \frac{1}{4} g_{MN}
\Psi_P^2$, the energy-momentum tensor (\ref{em-de})
simply reduces to \cite{hsy-jhep,ly-review}
\begin{equation}\label{de-1}
    T_{MN}^{(L)} \simeq - \frac{R}{32\pi G} g_{MN}
\end{equation}
where $R = \frac{1}{2 \lambda^2} \Big( \rho_M \rho_N + \Psi_M
\Psi_N \Big)g_{MN}$ is the Ricci-scalar of the metric (\ref{em-metric}). Note
that $\lambda^2 = \sqrt{-g}$ and $\rho_M = \partial_M \lambda^2$.
Since $\Psi_M$ is in some sense Hodge-dual to $\rho_M$, we expect
that their fluctuations will be of the same order with a
characteristic wavelength $L_H$. Therefore, by a simple dimensional
argument, we estimate that $R \sim \frac{1}{L^2_H}$. In the end, the
energy-momentum tensor (\ref{de-1}) proves to be\footnote{In the
Lorentzian signature, the sign of the Ricci scalar $R$ depends on
whether fluctuations  are spacelike $(R>0)$ or timelike $(R <0)$
\cite{hsy-jhep,ly-review}. In consequence the spacelike
perturbations act as a repulsive force whereas the timelike ones act
as an attractive force. When considering the fact that the
fluctuations in (\ref{em-de}) are random in nature and we are living
in (3+1) (macroscopic) dimensions, the ratio of the repulsive and
attractive components will end in $\frac{3}{4}: \frac{1}{4} = 75:
25$. Is it outrageous to conceive that this ratio curiously
coincides with the dark composition of our universe ?}
\begin{equation}\label{de-2}
    T_{MN}^{(L)} \sim - \frac{1}{L_P^2 L_H^2} g_{MN}.
\end{equation}

We will argue \cite{hsy-cc,hsy-ccc} that the weird energy
(\ref{de-2}) would be originated from ``vacuum fluctuations" with
the largest possible wavelength $L_H$ due to the UV/IR mixing
triggered by the noncommutativity of spacetime \cite{uvir}. Recall
that the vacuum energy $\rho_{\mathrm{vac}} \sim M_P^4$ does not
couple to gravity since it was used to create a flat spacetime in
our picture. Thus vacuum fluctuations over the vacuum (\ref{vacuum})
will be a leading contribution to the deformation of spacetime
curvature. So let us calculate the vacuum fluctuation energy by the
action (\ref{q-action}) with the largest possible wavelength $L_H$:
\begin{eqnarray} \label{de-3}
\rho &=& \rho_{\mathrm{vac}} + \delta \rho \nonumber \\
&=& \frac{1}{4 g_{YM}^2} \Big( B_{AB} - \widehat{F}_{AB}(x) \Big)^2
= \frac{1}{4 g_{YM}^2}  B^2_{AB} \Big(1 + \theta \widehat{F}(x)
\Big)^2 \nonumber \\
&\sim & M_P^4 \Big(1 + \frac{L_P^2}{L_H^2} \Big)^2 \sim M_P^4 +
\frac{1}{L_P^2L_H^2},
\end{eqnarray}
where it is natural to assess that $\widehat{F}(x) \sim
\frac{1}{L_H^2}$ for a dimensional reason. Note that the vacuum
fluctuation energy
\begin{equation} \label{dark-energy}
\delta \rho \sim \frac{1}{2 g_{YM}^2} B_{AB} \widehat{F}^{AB}(x)
\sim \frac{1}{L_P^2L_H^2}
\end{equation}
is a total derivative term and so a boundary term on a hypersurface
of radius $L_H$.\footnote{Unfortunately, the significance of
boundary terms such as (\ref{dark-energy}) due to the UV/IR mixing
was overlooked in \cite{hsy-jhep,ly-review}. This boundary
contribution caused by the UV/IR mixing could be consistent with the
holographic nature of dark energy.} If we assume $L_H$ to be the
size of cosmic horizon, the vacuum fluctuation energy
(\ref{dark-energy}) is in good agreement with the observed value of
current dark energy $\rho_{DE} \sim (10^{-3} \mathrm{eV})^4$
\cite{hsy-jhep}.

Although our conclusion about the dark energy may be too speculative
and so a full-fledged formulation is further required, we believe
that the emergent gravity from NC gauge fields has plenty of rooms
to explain the nature of dark energy and our underlying arguments
must be true even in a full-fledged theory.

\section{A novel unification in noncommutative spacetime}

It has been hoped that a physically viable theory of quantum gravity
would unify into a single consistent model all fundamental
interactions and describe all known observable interactions in the
universe, at both subatomic and astronomical scales. We have argued
that the gravitation can emerge from NC gauge fields and a
background independent quantum gravity can be defined by quantizing
spacetime itself. The upshot is that if gravity is emergent, then
the spacetime should be emergent too. If so, every structures
supported on the spacetime must also be emergent for an internal
consistency of the theory. Hence it should be natural that matter
fields as well as non-Abelian gauge fields for weak and strong
forces have to be emergent together with the spacetime. Thus an
urgent question is the following. How to define matter fields as
well as non-Abelian gauge fields describing quarks and leptons in
the context of emergent geometry ?

In order to figure out an underlying picture for emergent matters
\cite{hsy-jhep,ly-review}, it would be useful to start with the Feynman's observation about
the electrodynamics of charged particles \cite{feynman}. In 1948,
Feynman got a beautiful idea how to understand electrodynamics in
terms of symplectic geometry of particle phase space. Briefly
speaking, Feynman asks a question what is the most general form of
interactions consistent with particle dynamics defined in the
quantum phase space (\ref{qm}). Surprisingly he ends up with the
electromagnetic force. In other words, the electromagnetic force is
only a consistent interaction  with a quantum particle satisfying
the commutation relation (\ref{qm}). But the Feynman's observation
raises a curious question. We know that, beside the electromagnetic
force, there exist other interactions, weak and strong forces, in
Nature. Thus the problem is how to incorporate the weak and strong
forces together into the Feynman's scheme. Because he started with
only a few very natural axioms, there seems to be no room to relax
his postulates to include the weak and strong forces except
introducing extra dimensions. Remarkably it works with extra
dimensions !

To be more precise, consider a particle dynamics defined on
$\mathbf{R}^3 \times F$ with an internal space $F$ whose coordinates
are $\{ x^i: i =1,2,3 \} \in \mathbf{R}^3$ and $\{Q^I: I = 1,
\cdots, n^2 - 1 \} \in F$. The dynamics of the particle carrying an
internal charge in $F$ is defined by a symplectic structure on $T^*
\mathbf{R}^3 \times F$ whose commutation relations are given by \cite{feynman-lee,feynman-report}
\begin{eqnarray} \label{feynman-comm1}
 && [x^i, x^j] = 0, \qquad m[x^i, \dot{x}_j] = i \hbar \delta^i_j, \\
 \label{feynman-comm2}
 && [Q^I, Q^J] = i \hbar f^{IJK} Q^K, \\
 \label{feynman-comm3}
 && [x^i, Q^I] = 0.
\end{eqnarray}
The internal coordinates $Q^I$ satisfy $SU(n)$ algebra and so carry
their own Poisson structure inherited from the Lie algebra. One more
condition, the so-called Wong's equation \cite{wong}, is implemented
by
\begin{equation}\label{wong}
    \dot{Q}^I + \frac{1}{2} f^{IJK} (A_i^J Q^K \dot{x}_i + \dot{x}_i A_i^J Q^K) = 0
\end{equation}
to ensure that the internal charge $Q^I$ is parallel-transported
along the trajectory of a particle under the influence of
non-Abelian gauge fields $A_i(x,t) = A_i^I(x,t)  Q^I$ \cite{mongo}.
Actually, the Wong's equation is the Heisenberg equation
(\ref{h-eq}) for $Q^I$ with the Hamiltonian $\widehat{H} =
\frac{1}{2}( m \dot{x}^2_i + Q^I Q^I)$, i.e.,
\begin{equation}\label{wong-h}
    \dot{Q}^I = - \frac{i}{\hbar} [ Q^I, \widehat{H}].
\end{equation}
One can easily show it using the fact $m \dot{x}_i = p_i - A_i(x,t)$
and $[p_i, Q^I] = 0$.

If we repeat the Feynman's question, we can arrive at the conclusion
that the most general interaction of a quantum particle on
$\mathbf{R}^3$ carrying an internal charge $Q^I$ satisfying
$(\ref{wong})$ and the commutation relations
(\ref{feynman-comm1})-(\ref{feynman-comm3}) is a non-Abelian
interaction of $SU(n)$ gauge fields \cite{feynman-lee}. From our
perspective, we thus need extra dimensions with the Poisson
structure $F$ satisfying the above commutation relations to realize
leptons and quarks interacting with $SU(2)$ and $SU(3)$ gauge
fields. Wishfully, it will be more desirable to find a mechanism to
realize this structure together with the emergence of spacetime
geometry.

With this motivation, we consider a $U(N)$ gauge theory in four
dimensions whose action is given by
\begin{equation}\label{n=4}
S_{YM} = - \frac{1}{G_s} \int d^4 x \mathrm{Tr} \Big(
\frac{1}{4} F_{\mu\nu} F^{\mu\nu} + \frac{1}{2} D_\mu \Phi_a D^\mu \Phi_a
- \frac{1}{4} [\Phi_a, \Phi_b]^2 \Big)
\end{equation}
where $\Phi_a \; (a = 1,\cdots, 6)$ are adjoint scalar fields in
$U(N)$. For our purpose, we are interested in a large $N$ limit, in
particular, $N \to \infty$. The action (\ref{n=4}) is then exactly
the bosonic part of 4-dimensional $\mathcal{N} = 4$ supersymmetric
$U(N)$ Yang-Mills theory, which is the large $N$ gauge theory of the
AdS/CFT correspondence \cite{ads-cft1,ads-cft2,ads-cft3}. Suppose
that a vacuum of the theory (\ref{n=4}) is given by
\begin{equation}\label{n=4-vacuum}
    \langle \Phi_a \rangle_{\mathrm{vac}} = B_{ab} y^b, \qquad
    \langle A_\mu \rangle_{\mathrm{vac}} = 0
\end{equation}
where $B_{ab}$ is a constant matrix of rank 6. And assume that the
vacuum expectation values $y^a \in U(N \to \infty)$ satisfy the
algebra
\begin{equation}\label{nc6}
    [y^a, y^b] = i \theta^{ab} \mathbf{1}_{N \times N}
\end{equation}
where $\theta^{ab} = \big( \frac{1}{B} \big)^{ab}$. It is then
obvious that the vacuum (\ref{n=4-vacuum}) in the $N \to \infty$
limit is definitely a solution of the theory (\ref{n=4}) and the
vacuum algebra (\ref{nc6}) is familiar with the Heisenberg algebra
of NC space $\mathbf{R}_{\theta}^6$. Consequently the large-$N$
matrices on $\mathbf{R}^{3,1}$ in the action (\ref{n=4}) can be
mapped to NC fields in $C^\infty(\mathbf{R}^{3,1})
\otimes \mathcal{A}_\theta$.

Let us consider fluctuations $\widehat{A}_M (X) = (\widehat{A}_\mu,
\widehat{A}_a)(x,y), \; M = 0, 1, \cdots, 9$ of the large-$N$ matrices
in the action (\ref{n=4}) around the vacuum (\ref{n=4-vacuum})
\begin{equation}\label{matrix-fluc}
    D_\mu(x,y) = \partial_\mu - i \widehat{A}_\mu(x,y), \qquad
    \Phi_a(x,y) = B_{ab} y^b + \widehat{A}_a (x,y),
\end{equation}
where the fluctuations are assumed to also depend on the vacuum
moduli in (\ref{n=4-vacuum}). Therefore let us introduce
10-dimensional coordinates $X^M = (x^\mu, y^a)$ and 10-dimensional
connections defined by
\begin{eqnarray} \label{10d-conn}
D_M(X) &=& \partial_M - i \widehat{A}_M (X) \nonumber \\
  &\equiv& (D_\mu, D_a = - i \Phi_a) (x,y).
\end{eqnarray}
As a result, the large-$N$ matrices in the action (\ref{n=4}) are
now represented by their master fields which are higher-dimensional
NC $U(1)$ gauge fields in (\ref{10d-conn}) whose field strength is
given by
\begin{equation}\label{10-field}
    \widehat{F}_{MN} = \partial_M \widehat{A}_N - \partial_N
    \widehat{A}_M - i [\widehat{A}_M, \widehat{A}_N ]_\star.
\end{equation}
In the end, the 4-dimensional $U(N)$ Yang-Mills theory (\ref{n=4})
has been transformed into a 10-dimensional NC $U(1)$ gauge theory
and the action (\ref{n=4}) can be recast into the simple form
\cite{hsy-epj}
\begin{equation}\label{10-action}
    \widehat{S}_{10} = - \frac{1}{4 g_{YM}^2} \int d^{10} X \Big(
    \widehat{F}_{MN} - B_{MN} \Big)^2.
\end{equation}

To find a gravitational metric dual to the large-$N$ gauge theory
(\ref{n=4}) or, equivalently, to find an emergent metric determined
by the NC gauge theory (\ref{10-action}), we can apply the adjoint
operation (\ref{nc-deriv}) to the 10-dimensional NC gauge fields in
(\ref{10d-conn}) after switching the index $M \to A = 0, 1, \cdots,
9$:
\begin{eqnarray} \label{matrix-ad}
\widehat{V}_A [ \widehat{f}](X) &=& [D_A, \widehat{f}]_\star (x,y) \nonumber \\
  &=& V_A^M (x,y) \partial_M f(x,y) + \mathcal{O}(\theta^3)
\end{eqnarray}
for $\widehat{f}(x,y) \in C^\infty(\mathbf{R}^{3,1})
\otimes \mathcal{A}_\theta$. In the commutative limit, the
vector fields $V_A = V_A^M \partial_M \in \Gamma(TM)$ on a
10-dimensional Lorentzian manifold $M$ is given by
\begin{equation}\label{matrix-vector}
    V_A = (\partial_\mu + A_\mu^a \partial_a, D_a^b \partial_b)
\end{equation}
or their dual basis $V^A = V^A_M d X^M \in \Gamma(T^*M)$ is given by
\begin{equation}\label{10-dual}
    V^A = \Big( dx^\mu, V_b^a (dy^b - A^b_\mu dx^\mu) \Big),
\end{equation}
where $V_a^c D_c^b = \delta^b_a$ and
\begin{equation}\label{10-vector}
A_\mu^a \equiv - \theta^{ab} \frac{\partial
\widehat{A}_\mu}{\partial y^b}, \qquad
D_a^b \equiv \delta^b_a - \theta^{bc} \frac{\partial
\widehat{A}_a}{\partial y^c}.
\end{equation}
Hence the 10-dimensional geometry dual to the gauge theory
(\ref{n=4}) or (\ref{10-action}) can easily be determined by
\cite{hsy-epj,hsy-jhep}
\begin{eqnarray} \label{10-metric}
ds^2 &=& \lambda^2 \eta_{AB} V^A \otimes V^B \nonumber \\
  &=& \lambda^2 \Big(\eta_{\mu\nu} dx^\mu dx^\nu + \delta_{ab} V^a_c
  V^b_d (dy^c - \mathbf{A}^c) (dy^d - \mathbf{A}^d) \Big)
\end{eqnarray}
where $\mathbf{A}^a = A^a_\mu dx^\mu$ and the conformal factor is
defined by
\begin{equation}\label{10-conformal}
    \lambda^2 = \nu(V_0, V_1, \cdots, V_9)
\end{equation}
for a 10-dimensional volume form $\nu = d^4 x \wedge d^6 y$ or, more
generally, $\nu = d^4 x \wedge \nu_6$.

It has been known from the AdS/CFT duality
\cite{ads-cft1,ads-cft2,ads-cft3} that the large-$N$ gauge theory
(\ref{n=4}) is a nonperturbative formulation of type IIB string
theory on $AdS_5 \times \mathbf{S}^5$ background. We have verified
above that the 4-dimensional $U(N)$ gauge theory (\ref{n=4}) gives
rise to a 10-dimensional gravity with the metric (\ref{10-metric})
\cite{hsy-epj}. We see that the existence of nontrivial
gauge fields $A_\mu(x)$ causes the curving of the original flat
spacetime $\mathbf{R}^{3,1}$ and so the four-dimensional spacetime
also becomes dynamical together with an entirely emergent
6-dimensional space. Therefore, the large-$N$ gauge theory
(\ref{n=4}) almost provides a background independent description of
spacetime geometry except the original background $\mathbf{R}^{3,1}$
whose existence was {\it a priori} assumed at the outset. We confirm
again the important picture \cite{ly-review} that, in order to
describe a classical geometry from a background independent theory,
it is necessary to have a nontrivial vacuum defined by a ``coherent"
condensation of gauge fields, e.g., the vacuum defined by
(\ref{n=4-vacuum}).

A remarkable aspect of the large-$N$ gauge theory (\ref{n=4}) is
that it admits a rich variety of topological objects. So our
curiosity is what kind of geometry emerges from such a topological
object (according to the map (\ref{matrix-ad})) when the topological
solution has been defined by the gauge theory (\ref{n=4}) or
(\ref{10-action}) and what kind of object is materialized in
four-dimensional spacetime from the stable solution. We will assert
that consolidating some generic features of emergent geometry and
the Feynman's picture about the weak and strong forces leads to a
remarkable picture for what matter is. In particular, a matter field
such as leptons and quarks may simply arise as a stable localized
geometry in extra dimensions, which is a topological object in the
defining algebra (NC $\star$-algebra) of quantum gravity.

Consider a stable class of time-independent solutions in the action
(\ref{n=4}) satisfying the asymptotic boundary condition
(\ref{n=4-vacuum}). For such kind of solutions, we may forget about
time and work in the temporal gauge, $A_0 = 0$. Since the adjoint
scalar fields asymptotically approach the common limit
(\ref{n=4-vacuum}) (which does not depend on $x^i := \mathbf{x}$),
we can think of $\mathbf{R}^3$ as having the topology of a
three-sphere $\mathbf{S}^3 =
\mathbf{R}^3 \cup \{ \infty \}$, with the point at infinity included.
In particular, the matrices $\Phi_a(\mathbf{x})$ are nondegenerate
along $\mathbf{S}^3$ and so $\Phi_a$ defines a well-defined map
\begin{equation}\label{homotopy}
    \Phi_a: \mathbf{S}^3 \to \mathrm{GL}(N, \mathbf{C})
\end{equation}
from $\mathbf{S}^3$ to the group of nondegenerate complex $N \times
N$ matrices. If this map represents a nontrivial class in the third
homotopy group $\pi_3(\mathrm{GL}(N, \mathbf{C}))$, the solution
(\ref{homotopy}) will be stable under small perturbations, and the
corresponding nontrivial element of $\pi_3(\mathrm{GL}(N,
\mathbf{C}))$ represents a topological invariant \cite{horava}. In the stable
regime where $N > 3/2$, the homotopy groups of $\mathrm{GL}(N,
\mathbf{C})$ or $U(N)$ define a generalized cohomology theory, known
as K-theory $K(X)$ \cite{k-theory}. For example, for $X =
\mathbf{R}^{3,1}$, this group with compact support is given by \cite{horava}
\begin{equation}\label{k-group}
    K(\mathbf{R}^{3,1}) = \pi_{3}(\mathrm{GL}(N, \mathbf{C})) = \mathbf{Z}.
\end{equation}
Note that the map (\ref{homotopy}) is contractible to the group of
maps from $\mathbf{S}^3$ to $U(N)$.

We now come to the connection with K-theory, via the classic
Atiyah-Bott-Shapiro (ABS) construction \cite{abs} which relates the
Grothendieck groups of Clifford modules to the K-theory of spheres.
The ABS isomorphism relates complex and real Clifford algebras to
K-theory \cite{spin-geo}: Such a relation is somehow expected, given
that the periodicity of K-theory is similar to the periodicity of
Clifford algebras \cite{gtp-naka}. Note that the group $K(X)$ also
classifies D-branes in type II superstring theory on a manifold $X$
\cite{k1,k2,k3,k4}. In particular, the RR-charge of type IIB
D-branes is measured by the K-theory class of their transverse
space, so that $K(\mathbf{S}^{p}) = \pi_{p-1}(U(N))$ classifies
$(9-p)$-branes in type IIB string theory on flat $\mathbf{R}^{9,1}$
spacetime.

The ABS construction uses the gamma matrices of the Lorentz group
$SO(3,1)$ to construct an explicit generator of the K-theory group
(\ref{k-group}) \cite{spin-geo}. Let $S_\pm$ be two irreducible
spinor representations of $Spin(4)$ and define the gamma matrices
$\Gamma^\mu : S_+ \to S_-$ to satisfy the Dirac algebra $\{
\Gamma^\mu, \Gamma^\nu \} = 2 \eta^{\mu\nu}$. Let us also introduce
the Dirac operator $\mathcal{D}: \mathcal{H} \times S_+ \to
\mathcal{H} \times S_-$ such that
\begin{equation}\label{dirac-op}
    \mathcal{D} = \Gamma^\mu p_\mu + \cdots
\end{equation}
where $p_\mu = (\omega, \mathbf{p})$ is a four-momentum and the
abbreviation denotes possible higher order corrections in higher
energies. Here the Dirac operator (\ref{dirac-op}) is regarded as a
linear operator acting on a Hilbert space $\mathcal{H}$ as well as
the spinor vector space $S_\pm$. The Hilbert space $\mathcal{H}$
would be possibly much smaller than the primitive Fock space for the
Heisenberg algebra (\ref{nc6}) because the Dirac operator
(\ref{dirac-op}) actually acts on collective (coarse grained) modes
of the solution (\ref{homotopy}) \cite{horava}.

In order to construct stable topological objects that take values in
the K-theory (\ref{k-group}), it is natural to consider topological
solutions made out of $\Phi_a (x) \in U(N)$ according to the
homotopy map (\ref{homotopy}). As was shown before, the large-$N$
matrices in $U(N \to \infty)$ gauge theory can be described by NC
$U(1)$ gauge fields with the action (\ref{10-action}) in higher
dimensions. In particular, the adjoint scalar fields $\Phi_a (x)
\in U(N)$ are mapped to NC $U(1)$ gauge fields in extra dimensions
and obey the relation
\begin{equation}\label{adjoint-scalar}
    -i [\Phi_a, \Phi_b] = -B_{ab} + \widehat{F}_{ab}.
\end{equation}
Therefore, the topological solutions made out of $\Phi_a (x) \in
U(N)$ will be given by NC $U(1)$ instantons in four or six
dimensions. For instance, in four-dimensional subspace, one can
consider NC $U(1)$ instanton solutions given by
\cite{nc-inst,hsy-nci1,hsy-nci2}
\begin{equation}\label{nc-4inst}
    \widehat{F}_{ab} = \pm \frac{1}{2} {\varepsilon_{ab}}^{cd} \widehat{F}_{cd}
\end{equation}
where $a, \cdots, d = 1, \cdots, 4$ and, in six dimensions, one can
instead consider NC Hermitian $U(1)$ instantons defined by
\cite{our-future}
\begin{eqnarray} \label{nc-6inst1}
   && \widehat{F}_{ab} = \pm \frac{1}{4} {\varepsilon_{ab}}^{cdef}
    \widehat{F}_{cd} J_{ef}, \\
    \label{nc-6inst2}
    && J^{ab} \widehat{F}_{ab} = 0,
\end{eqnarray}
where $J_{ab}= \frac{1}{\kappa} B_{ab}$ is a nondegenerate
symplectic matrix defined by the vacuum (\ref{n=4-vacuum}). We
showed in section 2 that the NC $U(1)$ instantons in commutative
limit are equivalent to gravitational instantons which are
hyper-K\"ahler manifolds and also called Calabi-Yau 2-folds
\cite{hsy-epl,hsy-epj}. Similarly it can be shown \cite{our-future}
that the 6-dimensional NC Hermitian $U(1)$ instantons satisfying
(\ref{nc-6inst1}) and (\ref{nc-6inst2}) can be recast into
Calabi-Yau 3-folds using the vector fields defined by
(\ref{matrix-ad}). If we define a gravitational instanton as a
Ricci-flat, K\"ahler manifold, the Calabi-Yau 3-fold corresponds to
a 6-dimensional gravitational instanton.

\begin{figure}
\begin{center}
\includegraphics[width=0.5\textwidth]{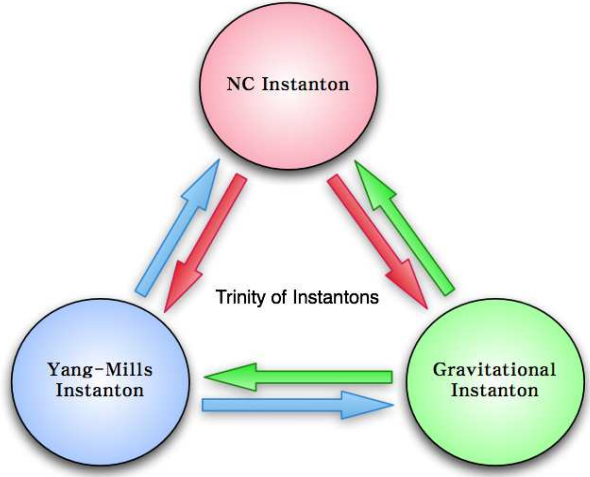}
\end{center}
\caption{\label{trinity}Trinity of instantons. (Image from \cite{loy})}
\end{figure}

It is well-known that gravity can be formulated as a gauge theory of
Lorentz group. In this gauge theory formulation, Calabi-Yau
$n$-folds correspond to $SU(n)$ Yang-Mills instantons in
$2n$-dimensions and the gauge group $SU(n)$ appears as the holonomy
group of a Calabi-Yau $n$-fold \cite{opy-jhep,yy}. Combining the
relationship between NC $U(1)$ instantons, $SU(n)$ Yang-Mills
instantons and Calabi-Yau $n$-folds altogether, we get the trinity
of instantons \cite{loy} depicted in Figure \ref{trinity}.

According to the above construction, the topological solution
(\ref{homotopy}) takes a value in the K-theory group (\ref{k-group})
and is realized as a stable geometry, i.e., a Calabi-Yau manifold,
in extra dimensions. And the ABS theorem says that the stable
topological solution is represented by the Dirac operator
(\ref{dirac-op}) on the spacetime $\mathbf{R}^{3,1}$. In other
words, the Calabi-Yau manifold constructed from a NC $U(1)$
instanton in extra dimensions will be realized as a four-dimensional
fermion $\chi(t, \mathbf{x})$ whose dispersion relation in low
energies is given by the relativistic Dirac equation
\begin{equation}\label{dirac-eq}
    i\Gamma^\mu \partial_\mu \chi + \cdots = 0
\end{equation}
as was already suggested by the Dirac operator (\ref{dirac-op}).
Although the emergence of 4-dimensional spinors from large-$N$
matrices or NC gauge fields is just a consequence of the fact that
the ABS construction uses the Clifford algebra to construct explicit
generators of $\pi_3(U(N))$, its physical origin is mysterious and
difficult to understand.

Recall that the Weyl spinor in (\ref{dirac-eq}) is originated from
NC $U(1)$ instantons in extra $2n$-dimensions which are also
realized as Calabi-Yau $n$-folds with $SU(n)$ holonomy. Therefore
the 4-dimensional spinor $\chi$ should be charged under the $SU(n)
\subset SO(2n)$ gauge group and so can couple to $SU(n)$ gauge fields
in four dimensions. If so, our last question is how non-Abelian
gauge fields $A_\mu^I(x) \in SU(n)$ on $\mathbf{R}^{3,1}$ can arise
from the $U(1)$ gauge fields on $\mathbf{R}^{3,1} \times
\mathbf{R}_\theta^6$.

To recapitulate, the $U(N \to \infty)$ gauge theory (\ref{n=4}) in
the Moyal background (\ref{n=4-vacuum}) has been mapped to the
10-dimensional NC $U(1)$ gauge theory (\ref{10-action}) defined on
the space $\mathbf{R}^{3,1} \times \mathbf{R}_\theta^6$. Then the
K-theory (\ref{k-group}) for any sufficiently large $N$ can be
identified with the K-theory $K(\mathcal{A}_\theta)$ for the NC
$\star$-algebra $\mathcal{A}_\theta$. But, if we consider low-energy
excitations around the solution (\ref{homotopy}) whose K-theory
class is given by $K(\mathcal{A}_\theta)$ and that would be a
sufficiently localized state described by a compact (bounded
self-adjoint) operator in $\mathcal{A}_\theta$, it will not
appreciably disturb the ambient gravitational field. This means
\cite{hsy-jhep} that we may reduce the problem to a quantum particle dynamics on
$\mathbf{R}^{3,1} \times F$ where $F$ is an internal space
describing collective modes of the solution (\ref{homotopy}). It is
natural to identify the coordinate of $F$ with an internal charge
carried by the Weyl fermion $\chi$ in (\ref{dirac-eq}). We observed
above that the (collective) coordinates of $F$ will take values in
the $SU(n)$ Lie algebra such as the isospins or colors and will be
denoted by $Q^I \; (I = 1, \cdots, n^2 -1)$.

It is useful to remember that the vacuum algebra (\ref{nc6}) is
nothing but the $3$-dimensional quantum harmonic oscillators with
the Heisenberg algebra
\begin{equation}\label{3-ho}
[y^a, y^b] = i\theta^{ab} \qquad \Leftrightarrow \qquad    [a_i,
a_j^\dagger] = \delta_{ij}
\end{equation}
where $a,b = 1, \cdots, 6$ and $i,j =1,2,3$. There is a well-known
fact that the $n$-dimensional harmonic oscillator can realize an
$SU(n)$ symmetry and the generators of the $SU(n)$ Lie algebra are
given by
\begin{equation}\label{schwinger}
    Q^I = \sum_{i,j} a_i^\dagger T^I_{ij} a_j
\end{equation}
where $T^I$ are constant $n \times n$ matrices satisfying the
$SU(n)$ Lie algebra $[T^I, T^J] = i \hbar f^{IJK} T^K$. It is easy
to check that the Schwinger representation (\ref{schwinger})
satisfies the commutation relations in
(\ref{feynman-comm1})-(\ref{feynman-comm3}). As was reasoned above,
the $SU(n)$ generators in (\ref{schwinger}) can be regarded as
low-energy collective modes (or order parameters) in the vicinity of
the solution (\ref{homotopy}).

Since the chiral fermion $\chi$ in (\ref{dirac-eq}) is charged under
the $SU(n)$ symmetry whose generators are given by
(\ref{schwinger}), it can interact with four-dimensional $SU(n)$
gauge fields $A_\mu^I(x)$. Let $\rho(\mathcal{H})$ be a
representation of the Lie algebra (\ref{schwinger}). We will take an
$n$-dimensional representation in $\mathcal{H} = L^2(\mathbf{C}^n)$
which is much smaller than the original Fock space of (\ref{3-ho}).
Since we are considering a low-energy limit where gravitational
back-reactions are ignored, it will be reasonable to take only the
lowest modes of NC $U(1)$ gauge fields $\widehat{A}_\mu(x,y) \in
C^\infty(\mathbf{R}^{3,1}) \times
\mathcal{A}_\theta$ as a low-energy approximation. So we will expand
the $U(1)$ gauge fields $\widehat{A}_\mu(x,y)$ in
(\ref{matrix-fluc}) with the $SU(n)$ basis in (\ref{schwinger})
\begin{equation}\label{expansion-u1}
\widehat{A}_\mu(x,y) = A_\mu(x) + A_\mu^I(x) Q^I +  A_\mu^{IJ}(x)
Q^I Q^J + \cdots
\end{equation}
where it is assumed that each term in (\ref{expansion-u1}) belongs
to an irreducible representation of $\rho(\mathcal{H})$. Through the
expansion (\ref{expansion-u1}), we get $SU(n)$ gauge fields
$A_\mu^I(x)$ as well as ordinary $U(1)$ gauge fields $A_\mu(x)$ as
low lying excitations \cite{hsy-jhep}.

The coarse-grained fermion $\chi$ in (\ref{dirac-eq}) behaves like a
stable relativistic particle in the spacetime $\mathbf{R}^{3,1}$.
Hence, when the particle moves along $\mathbf{R}^{3}$, there will be
bosonic excitations arising from changing the position in
$\mathbf{R}^{3}$ of the internal charge $F$ according to the
relation $m \dot{x}_i = p_i - A_i(\mathbf{x}, t)$ and the Wong's
equation (\ref{wong-h}). That is, we can think of the Dirac operator
(\ref{dirac-op}) as an operator $\mathcal{H} \times S_+ \to
\mathcal{H} \times S_-$ where $\mathcal{H} = L^2(\mathbf{C}^n)$ and
introduce a minimal coupling with the $U(1)$ and $SU(n)$ gauge
fields in (\ref{expansion-u1}) by the replacement $p_\mu \to p_\mu -
eA_\mu(x) - A_\mu^I (x) Q^I$. Then the Dirac equation
(\ref{dirac-eq}) becomes
\begin{equation}\label{dirac-coupling}
    i \Gamma^\mu \Big(\partial_\mu - ie A_\mu - iA_\mu^I (x) Q^I \Big) \chi +
    \cdots = 0.
\end{equation}
Here we see that the chiral fermion $\chi$ in the homotopy class
$\pi_3(U(N))$ is in the fundamental representation of $SU(n)$. As a
result, the spinor in (\ref{dirac-coupling}) can be identified with
a quark, an $SU(3)$ multiplet of chiral Weyl fermions interacting
with gluons $A_\mu^I(x) \; (I=1, \cdots, 8)$ for $n=3$ and with a
lepton, an $SU(2)$ doublet of chiral Weyl fermions interacting with
isospin gauge fields $A_\mu^I(x) \; (I=1, \cdots, 3)$ for $n=2$
\cite{hsy-jhep,ly-review}.

We want to point out that the emergent matters from stable
geometries in extra dimensions are consistent with the Calabi-Yau
compactification in string theory. In string theory, a Calabi-Yau
manifold serves as an internal geometry of string theory with 6
extra dimensions and their shapes and topology determine a detailed
structure of the multiplets for elementary particles and gauge
fields through the compactification, which leads to a low-energy
phenomenology in four dimensions. A very similar picture seems to be
also realized in the context of emergent geometry via the ABS
theorem and the trinity of instantons illustrated in Figure
\ref{trinity}.

To conclude, we have observed that the theory (\ref{n=4}) allows
topologically stable solutions as long as the homotopy group
(\ref{k-group}) is nontrivial. Remarkably, a matter field such as
leptons and quarks simply arises from such a stable solution and
non-Abelian gauge fields correspond to collective zero-modes of the
stable localized solution (\ref{homotopy}). Although the solution
(\ref{homotopy}) is interpreted as particles and gauge fields
ignoring their gravitational effects, we have to recall that it is a
stable excitation over the vacuum (\ref{n=4-vacuum}) and so
originally a part of spacetime geometry according to the map
(\ref{matrix-ad}). Consequently, we get a remarkable picture, if
any, that matter fields such as leptons and quarks simply arise as a
stable localized geometry, which is a topological object in the
defining algebra (NC $\star$-algebra) of quantum gravity. This
approach for quantum gravity thus allows a novel unification where
spacetime as well as matter fields is equally emergent from a
universal vacuum of quantum gravity \cite{eg-review3}. We believe
that such an elegant unification of geometry and matters is a unique
feature realized only in the background independent formulation of
quantum gravity.

\ack
We are grateful to Jungjai Lee and John J. Oh for helpful
discussions, collaborations and encouragements over the years. This
research was supported by Basic Science Research Program through the
National Research Foundation of Korea (NRF) funded by the Ministry
of Education, Science and Technology (2011-0010597) and by the
RP-Grant 2010 of Ewha Womans University.

\end{document}